\documentstyle[11pt,aaspp]{article}

\begin {document}

\title{A Multiwavelength Study of Star Formation in the L1495E Cloud in Taurus}
\author{Karen M. Strom\altaffilmark{1} and Stephen E. Strom\altaffilmark{1}}
\affil{Five College Astronomy Department, Graduate Research Center, 517-G,\\
University of Massachusetts, Amherst, MA 01003\\
Electronic Mail: kstrom@hanksville.phast.umass.edu;
sstrom@donald.phast.umass.edu}

\altaffiltext{1}{Visiting Astronomer, Kitt Peak National Observatory,
National Optical Astronomy Observatory, which is operated by the Association
of Universities for Research in Astronomy, Inc. (AURA) under cooperative
agreement with the National Science Foundation.}

\begin{abstract}

We have carried out a deep (t=30000s) x-ray search of the eastern portion
of the L1495 cloud centered on the well known weak line T Tauri star (WTTS)
V410 Tau using the ROSAT PSPC. This deep exposure enabled a search for
candidate pre-main
sequence (PMS) objects in this cloud to a limit $\sim 20$ times more
sensitive than that typical of the fields examined with the
{\em Einstein} searches. Despite assertions that the PMS population in
Taurus-Auriga is nearly completely known, this x-ray survey revealed 8
new PMS objects in a region 50\arcmin$\:$ in diameter, as compared to a
previously known stellar population of 12 objects, including deeply embedded
IRAS sources.

Spectroscopic and photometric observations enable us to place these
objects in the HR Diagram. The newly discovered objects are
predominantly stars of spectral type M0 and later, and a large fraction (6/8)
appear to be surrounded by circumstellar accretion disks as judged by
their infrared excess and H$\alpha$ emission. We combined
the data for these x-ray discovered objects with extant and new data
for the previously identified PMS stars in this region to examine the
history of star formation and the frequency distribution of stellar
masses in this cloud.

If the ``post ROSAT'' population is either complete or
representative, we conclude (1) that star formation in L1495 East took place
$\sim1\times 10^6$ yrs ago and that the spread in ages is small; (2)
the frequency distribution of masses, N(M), in this apparently coeval
group appears to peak near $\log M = -0.5$ (using masses derived from the
recently published PMS tracks of D'Antona \& Mazzitelli (1993) and Swenson et
al.\
(1993)) and to
decline toward lower masses. The derived N(log M) for L1495E compares well with
the IMF derived from studies of stars in the solar
neighborhood, a result which suggests that the Taurus-Auriga clouds are
currently
producing stars whose mass spectrum approximates the time/space averaged
IMF for the solar neighborhood.

\end{abstract}

\section{Introduction}

X-ray searches in regions of recent star formation have proven
efficacious not only in recovering previously known solar-type PMS
objects (the classical T Tauri stars or CTTS), but in locating a
class of objects of which few were previously identified, the weak line
T Tauri stars  (WTTS) which lack the emission phenomena (H$\alpha$
emission; metallic line emission; optical excess (veiling) and infrared
excess emission) which
characterize CTTS. We report in this contribution the results of a deep
search for x-ray sources in the vicinity of an actively star forming
cloud within the Taurus-Auriga cloud complex: L1495 E(ast). Our ROSAT
PSPC image is the result of a total integration time of $\sim 33000$s
and is centered on V410 Tau. We had originally hoped to use these data
to search for periodic variations in x-ray flux from V410 Tau. However
the exposure did not span a large enough fraction of the 1.88 day
rotation period to provide a definite result. The long effective integration
time did permit a search for sources with x-ray luminosities as faint as
$3.0\times10^{28}\,$ ergs/s (unreddened), $\sim 20\:$ times deeper than
any previous x-ray exposure in this region. As a result we located
a number of additional candidate PMS
stars (8 in number, compared to a previously identified PMS population
of 12 objects in the same area). The goals of this study are (1) to
locate optical counterparts for the newly identified sources; (2) to
locate these objects in the HR Diagram using red spectra and optical and
near infrared photometry; (3) to determine masses, ages and disk
frequency for these objects; and (4) to construct an IMF for this
population.

\section{Cloud Environment and Previous Studies of the Stellar Population}

Large scale studies of the L1495 molecular cloud have been published by
Duvert, Cernicharo \& Baudry (1986) and Fukui et al.\ (1992). These maps
were made with radio telescopes of 2.5m and 4m diameter and therefore of
relatively
low spatial resolution ($\theta \sim 2\farcm 5$). However both maps show that
the greatest
concentration of molecular material is centered on the region containing
V410 Tau, V892 Tau, DD Tau and CZ Tau, the region upon which our ROSAT PSPC
image
is centered. Unfortunately, the region has not yet been studied either
at high spatial resolution or in molecules sensitive to
high densities (e.g. NH$_3$ ). Most molecular studies have been concentrated in
the less dense western part of
this cloud.  In view of the youth of this region and the degree of star
forming activity, it would be of great interest to pursue such studies.

Recent observations with the Nagoya University radio telescope, centered on
known PMS objects
show that the velocity dispersion as measured by line widths in this part of
L1495
is larger than that seen in any other part of the Taurus-Auriga clouds,
$\geq$3 km/s. The velocity dispersion drops considerably (to 1-1.5 km/s) in the
outer parts of the cloud. In the western star forming center of L1495
(centered on V773 Tau) the
velocity dispersion is $\sim 2$ km/s. Duvert et al.\ (1986) attributed the
measured velocity dispersion to multiple components with a cloud collision
possibly underway. However, while some of the line profiles appear to
have weak additional components, the line profiles at cloud center
appear to be smooth gaussians with FWHM of $\geq3$ km/s.
The mass of L1495E, within the region covered by the ROSAT PSPC image,
lies between 180 $M_\odot$ as determined from the Nagoya C$^{18}$O map and
230 $M_\odot$ as determined from the $^{13}$CO map (Ohnishi \& Mizuno 1993)

Within a circle of 1\deg$\,$ radius surrounding V410 Tau there are 19
known PMS objects; 12 of these lie within a circle of
25\arcmin $\,$ radius centered in the ROSAT field. There is one Herbig Ae/Be
star, V892 Tau (Elias 1),
centrally located in the cloud; also located in the central region are 5 WTTS,
3
CTTS and 3 heavily embedded Class I sources (Lada 1987), one of which is
known to drive a jet seen in H$\alpha$ and [\ion{S}{2}] (Strom et al.\
1986, Goodrich 1993). In the outer region, 30\arcmin~$<$~r~$<$~55\arcmin,
there are 4 CTTS and 3 WTTS. While M type stars are
known among this population, they are all {\em early} M stars (spectral types
M2 or earlier). This region of
the Taurus clouds thus exhibits the spectral type distribution characteristic
of the rest of the Taurus clouds which show an apparent deficiency of late M
stars, spectral types later than M2 (Cohen \& Kuhi
1979), and therefore an apparent deficiency of stars with masses $M <
0.2 M_\odot$.

\section{Observations}

The x-ray observation of the L1495E cloud was obtained in pointing mode with
the ROSAT
telescope using the position sensitive proportional counter (PSPC) ,
between $4^h 09^m$ on 1991 Mar. 4 UT and $23^h 25^m$ on 1991 Aug. 2 UT.
The total observation time over this period was 32986s. A description of the
satellite, the
telescope and the detector are given by Tr\"{u}mper (1983) and
Pfeffermann et al.\ (1986).
Because a portion of the data included in the standard analysis (SASS, version
5-6)
contained observations taken during a period
when the satellite was rolling (as indicated by the ``housekeeping'' data)
the observation was rereduced excluding data taken during this period;
the effective observation time is 25757s.
The data were rereduced, excluding the portion affected by satellite
roll, and analyzed using the PROS package (versions 2.1 and 2.2) under
IRAF\footnotemark[2].
\footnotetext[2] {IRAF is distributed by the National Optical Astronomy
Observatories, which is operated by the Association of Universities for
Research in Astronomy, Inc. (AURA) under cooperative agreement with the
National Science Foundation.}

The optical photometry at R and I was obtained on the night of 1992 Dec 11 UT
with a Tektronix 2048x2048 CCD
mounted in a universal CCD dewar on the Kitt Peak National Observatory 0.9m
telescope by Patrick Hartigan. The plate
scale for the camera/telescope combination used for this observation was
0\farcs775/px. A series of exposure times ranging from 10s through
300s was used for observations at I in order to obtain unsaturated images
of the bright stars. A single exposure at R of 30s duration was made. Only 2
stars suffered from saturated images in this exposure. Since the field was
not crowded, the photometry was reduced using the DIGIPHOT package in IRAF
using the new Landolt
standard fields (Landolt 1992) and the PHOTCAL routine (Davis 1993).

The near infrared photometry was obtained during 4 different runs due to
the convolved effects of the small size of infrared arrays and the terrible
weather in Arizona during the winter of 1993. The largest batch of near
infrared observations was taken by Ronald Probst on the night of 1993 Jan
12 UT using the SQIID photometer on the 1.3m telescope at KPNO.
Simultaneous images in J, H, and K were obtained using
256 x 256 PtSi arrays over a field of dimension $5 \farcm 4$ x $5 \farcm 4$;
the size of each pixel is 1\farcs36. Each image is the sum of 2
``dithered'' images, offset by $\sim20\arcsec \:$ with an integration time of
180s apiece. One additional field was obtained by Lynne Hillenbrand and
the authors on 1992 Dec 27 UT, again with SQIID on the 1.3m telescope at KPNO.
The field centered on V892 Tau (Elias 1) was observed on 1992 Jan 27 UT with
this same instrumentation by the authors. These data were all reduced using
the SQIID package under IRAF. A few additional single channel photometer
measurements collected for other programs at San Pedro Martir are also
included.

One additional field, as well as 2 previously observed fields, were observed by
Michael Meyer and Patricia Knesek
on 1993 March 11 UT at the 2.4m Hiltner telescope at the MDM observatory on
Kitt Peak using the NicMass
camera, equipped with a 256 x 256 HgCdTe NICMOS array having a field of
dimension $4 \farcm 25$ x
$4 \farcm 25$, and a plate scale of $0 \farcs 81$/px. A series of four 3s
integrations in each filter was followed by a series of four 10s
integrations. These data were also reduced using the SQIID package under
IRAF.

Spectra were obtained with HYDRA, the fiber optic coupled
spectrograph on the 4m telescope at KPNO. We used the red fiber cable in
conjunction with the BL181 grating (316 l/mm) set at a central
wavelength of 7600\AA, and a Tektronix 2048x2048 CCD
to obtain spectra with a usable range from $5500\AA \,$ to $9200\AA$; the
effective resolution was $\sim$3 pixels or 6\AA. The spectra were then
reduced using the DOHYDRA package within IRAF.
Further analysis was done using the ONEDSPEC package. Spectral standards were
established using the main sequence proper motion members of Praesepe
(Jones \& Stauffer 1991) for spectral types between F6 and M4. These standards
were
supplemented by the spectra of
Kirkpatrick, Henry \& McCarthy (1991) for the late M stars.

\section{Identification of Optical Counterparts to ROSAT Sources}

Our deep ROSAT PSPC exposure centered on V410~Tau allows us to assess the
completeness of the known pre-main sequence stellar
population in this nearby active star formation region, of which
only the periphery was observed with the {\it Einstein}
satellite.  The pre-ROSAT surveys do not provide the basis for evaluating
the number of very low mass stars ($M \leq 0.3 M_\odot$) in this
region, leaving the question of the population of the low
mass end of the IMF unaddressed.
Some information is provided by extant H$\alpha$ surveys
and the LkCa survey.
This part of the Taurus clouds was covered by the Herbig, Vrba \&
Rydgren (1986) survey for stars
exhibiting Ca~II H \& K emission and three such stars (LkCa 4, 5, \& 7)
were found within the
field of this ROSAT observation. These stars exhibit H$\alpha$ emission
and strong \ion{Li}{1} $\lambda 6708\AA$ absorption (Strom et al.\ 1989a,
Walter et al. 1988)
and are clearly
pre-main sequence objects. However their H$\alpha$ emission equivalent widths
are
too small to allow detection by H$\alpha$ objective prism surveys.
Unfortunately, both the
H$\alpha$ and \ion{Ca}{2} emission line surveys reach only the brighter stars
in the Taurus clouds (R $\leq$ 13), and are incomplete for
fainter stars, usually locating only those faint stars with unusually strong
H$\alpha$
emission.

Our deep ROSAT PSPC image of the V410 Tau region should enable a deep
search for faint pre-main sequence stars based on
the elevated x-ray luminosity characteristic of PMS stars. Relative to
main sequence stars of the same spectral type, the x-ray
luminosities of PMS stars may be $10^3$ times greater (Feigelson 1986,
Strom et al.\ 1990). Both CTTS and WTTS are
strong x-ray emitters, with no significant difference in the x-ray
properties of the two groups (Feigelson \& Kriss, Strom et al.\ 1990,
Gauvin \& Strom 1992).
Our observation should allow us to detect lightly obscured stars at the
distance
of the Taurus clouds (160 pc) having x-ray luminosities a
factor of 20 lower than those found with the {\it Einstein}
observations.

Unfortunately the sensitivity of a ROSAT PSPC observation is not uniform
across the field.
The point spread function of the ROSAT PSPC is a strong nonlinear function of
the
distance from field center, varying from FWHM $\sim25\arcsec \,$ at field
center to $1\arcmin \,$
at $\sim 30\arcmin \:$ from field center to $4\arcmin \;$ at the edge of the
$2\deg \,$ field; therefore the
sensitivity decreases rapidly
outside the central 25\arcmin$\,$ of the field. All but 3 (CK Tau 1,
PSC04154+2823, PSC 04158+2805) of the 12 previously identified
pre-main sequence stars (Herbig \& Bell 1988) within the central field of
radius $25\arcmin \,$
 were
easily detected. There is a faint extension on the x-ray image of Hubble 4 that
may be attributable to CK Tau 1. In the outer annulus $(25\arcmin \leq r \leq
55\arcmin)\;$ 6 more
previously identified members (RY Tau, BP Tau, HDE 283572, FO Tau, LkCa
4, LkCa 7) of the Taurus cloud population are found.
One of the previously known PMS stars,
FO Tau, is not detected in this exposure.  It is located at a distance
$51\farcm 8 \;$ from the field center. While other PMS stars were detected at
a distance from field center almost this great,
(notably RY Tau and HDE 283572 at $r = 46\farcm2$),
the much lower luminosity of FO Tau apparently precludes its detection.
In addition to the previously known members of the PMS population, eight
additional x-ray detections ($> 5 \sigma$) appear in the central region of our
exposure,
which is centered on the densest part of the
dark cloud. The positions of the previously known PMS stars are offset
from the x-ray positions derived from the standard analysis by $\Delta \alpha =
-3\farcs8 \pm 1\farcs0$ and
$\Delta \delta = 8\farcs0 \pm 2\farcs5$. Using these offsets as a
template, we were able
to identify optical counterparts for each of the new x-ray sources on our deep
I band CCD frame. Within the
positional error given by the local x-ray point spread function
($\Delta\theta \leq 10\arcsec$), 6
sources (V410 x-ray 1, 2, 3, 4, 6, \& 7) appear to have unique optical
counterparts while only two
sources (V410 x-ray 5 \& 8) within the central region of the field had more
than one optical
object as a possible identification. The object farthest from field center
(V410 x-ray 8, r
= 18\farcm 5) is located near the cloud periphery, hence complicating the
identification owing to increasing contamination by background field
stars. In this case
we examined 6 possible identifications. Although most of these 6
candidates are much too far from
the x-ray position, we felt that the possibility that the brightest star
of the 6,
x-ray 8a, was the true x-ray source should be explored.
The positional information
for the sources detected in our ROSAT
observation, as well as for the other PMS objects in the field, are presented
in Table 1. The table gives the object name, the
optical position, the x-ray position, the difference between the x-ray and
optical positions in the sense x-ray minus optical, and the radial distance
from field
center. A finding chart is given in Figure 1 (Plate XX).

\subsection{X-ray Properties of the Sources}

For each detected x-ray photon, the time of arrival, position and energy are
recorded.
This allows us to evaluate the variability of the source and to fit its
spectral energy distribution with models which take into account
both extinction from the intervening interstellar medium and the source
temperature.
Assuming that the x-ray emission from each of these sources arises in a stellar
corona, we model
the observed spectrum with a Raymond-Smith (Raymond \& Smith 1977) thermal
plasma, assuming solar
abundances. We then solved for the
temperature of the plasma and the column density of Hydrogen atoms along
the line of sight assuming the Morrison-McCammon (Morrison \& McCammon 1983)
interstellar
x-ray absorption model. For the brighter x-ray sources, the signal to noise
was sufficiently high to note strong systematic behavior in the fit residuals.
Where appropriate, these
sources were then refit with a two temperature model (Schrijver, Lemen \&
Mewe 1989)
where the line of sight extinction was held constant but two components
of different temperature were included in the fit. These models provided
substantially better fits (the $\chi ^2$ values were improved by factors of
5 to 35) to the x-ray spectra of the brighter sources.
Single component models were used for the fainter sources.
The results of these fits for all stars are given in Table 2.
Examples of the fits obtained for two high signal to noise sources (V410 Tau
and Hubble 4) with different line of sight extinctions are shown in Figure 2.
Also shown are the $\chi^2$ plots in the
temperature-$\log N_H$ plane. These plots
demonstrate the sensitivity of the fits to line of sight extinction,
even when the count rate is high.
Using these fits, we
are then able to calculate extinction corrected x-ray luminosities
(0.2-2.4 KeV) for these sources. When line of sight extinction is
sufficiently high that few low energy photons are detected, the error in
the temperature determination increases, sometimes substantially.
The two temperature model serves only as an approximation to the more likely
situation of a continuous temperature distribution. Therefore we feel that the
presentation of formal error bars for the temperature and emission measure
values would not be physically meaningful. Higher spectral resolution is
necessary in order to make more physically meaningful measurements of the
temperature distribution found in the region where the x-ray emission is
formed.
For sources with too few counts to yield a reliable fit, a model with $\log
N_H = 21$ and $T = 1$ KeV was fit in order to allow an estimate of the
x-ray luminosity.

We can compare the distribution of x-ray luminosities of the newly
discovered x-ray objects with those of the previously identified PMS
population of L1495E. The distributions of the 2 populations are shown
in Figure 3. It is obvious that, while the {\em range} in x-ray luminosity for
the 2 groups is similar, the {\em distributions} are vastly different. The
median x-ray luminosity for the previously identified population is
$\log L_x =$ 30.26; the median x-ray luminosity for the newly identified
population is $\log L_x =$ 29.00. Therefore, by substantially increasing
our sensitivity to low luminosity sources, we have demonstrated that {\em PMS}
stars may have
x-ray luminosities at least as low as those found for Pleiades {\em main
sequence} stars of solar type (Stauffer et al. 1993).

While it was hoped that this observation would cover a full rotation
period of V410 Tau,
scheduling constraints led to the coverage of only $\sim\onehalf $ of the
period of this star. V410 Tau shows clear evidence of high amplitude variation.
We were unable
to comment definitively on the relationship between the x-ray light curve and
that seen in the optical. However we are able to extract variability
data for each of our detected objects on shorter timescales because
the arrival time of each detected photon is recorded as part of the
observation. The
Kolgomorov-Smirnov and Cramer-von Mises tests for constancy were performed on
the time sorted event files for the brighter sources in this field.
Nine of the 12 sources bright enough to yield reliable results
demonstrate clear variability.
These results are also tabulated in Table 2 where we list the probability of
variability found for the sources examined as well as the ratio of the maximum
count rate to the minimum count rate seen in an 860s bin.
The large number of x-ray variables is not surprising in light of the
extreme variability already noted for some of the PMS sources in the {\it
Einstein}
observations (Montmerle et al.\ 1983, Feigelson \& DeCampli 1981).
None of the 8 newly discovered x-ray sources in L1495E was
sufficiently bright to enable reliable variability indications.

At least one, and possibly 2, x-ray flares
occurred on PMS stars within the field during the observation
period (Figure 4). DD~Tau
clearly flared during the second half of the observation period following a
long period
of very low x-ray flux. Unfortunately, owing to a gap in the observation,
we are unable to determine the rise time of the flare.
The emission at the peak of the flare was a factor of 15 higher than that
seen prior to the flare. The e-folding time for the decay of the flare is
$\sim$1 hour;
however the star remained a factor of 2 above the flux level
of its previous quiesent stage more than 5 hours after the flare began,
at the end of the observation.
It is possible that a less extreme event was seen
on HDE 283572. Here the rise time is $\sim 1.4$ hours and the decay time is
$\sim 6$ hours. The peak flux is $\sim 6$ times the pre-event flux.
Another interpretation of the HDE 283572 observation would be that we are
seeing variability
associated with the rotation of an active region on the surface of the
star. The rotation period of this star is $\sim 37$ hours (1.55d) (Walter
et al.\ 1988); therefore our observation covered 0.65 periods. Lacking
more data, we are unable to determine which interpretation is more
likely.

\section{The Optical and Near Infrared Photometry and Spectroscopy for
the Candidate Objects}

We have obtained new Cousins R and I band CCD observations of the
central region of the ROSAT field, as well as new near infrared (J, H,
and K) observations of the x-ray sources in the same region. These data
are presented in Table 3. When multiple observations have been made, the
epoch is also given.

We were also able to obtain classification
spectra in the red region for most of the possible x-ray source
identifications (see Figure 1) with the HYDRA fiber optic fed
spectrograph. Because of restrictions imposed by the
placement of the optical fibers in the focal plane of the telescope,
when a choice had to be made between two or
more objects, the more likely identification (the brightest star nearest
the x-ray position) was chosen. For this
reason, no spectra were obtained of V410 x-ray 5b, V410 x-ray 8b and c.
Although
fibers were placed on V410 x-ray 2 and 4, no usable spectra were obtained owing
to the faintness of the stars. No spectrum was obtained for DD~Tau
on this run. The spectral type for this star was taken from Hartigan,
Strom \& Strom (1994). The spectral type for CZ Tau is a compromise
since there is a definite change of spectral type with wavelength of at
least 2 spectral subclasses, with
later spectral types being found at longer wavelengths. This does not
appear to be due to a variable veiling contribution, but is perhaps due
to contamination of the spectrum with light from the subarcsecond
companion found by Leinert et al.\ (1992).

Observations of the Praesepe main sequence stars were used as spectral
standards with which to classify our spectra of the x-ray sources in
L1495E. The spectral types obtained are also given in Table 3; the
spectral types should be good to 1 spectral subclass earlier that M2 and
\onehalf$\,$ a spectral subclass for stars later than M2. While the
resolution of the spectra are too low to enable measurement of the strength of
the Li
$\lambda 6708${\AA} line, it was possible to measure the equivalent width of
the H$\alpha$ emission line, thus providing another criterion for including the
star as a candidate member of the PMS
population. These values are presented in Table 3 as well.

\subsection{Anonymous Stars}

Because there were 97 fibers available for placement upon stars or sky
positions when using the HYDRA spectrograph, we chose to place some of
these fibers upon anonymous stars that fell within the area of the L1495E
cloud. We obtained usable spectra for 5 of these stars, optical photometry
for these 5, and, coincidentally, near infrared photometry for 4 of the
stars. This data is also given in Table 3.

Other anonymous stars appeared in our infrared imaging frames. Data for
these stars is also given in the Appendix.

\section{Stellar Properties}

On the basis of the data tabulated in Table 3, we can assess the
likelihood that we have properly identified the stellar x-ray sources
and compare the distribution of the spectral types of the newly
identified stars with that of the previously identified population.
We must first evaluate the extinction suffered by each star.
The spectral type and observed R-I color give
a color excess for each star, E(R-I), which, when combined with a
standard reddening law, yields the extinction. These values are also
tabulated in Table 3.

Among the candidate objects for identification with the x-ray sources,
only the spectrum for V410 x-ray 8a clearly disqualifies it as a member
of the cloud population. The spectral
type of V410 x-ray 8a is M0-2 III. This spectral type combined with the
available photometry indicates substantial extinction. Since the object lies
near the cloud edge, it  is likely that we are viewing the star through a
substantial path length of the L1495 cloud;
the luminosity classification of this
spectrum clearly identifies the star as a background giant.
The positional discrepancy with the x-ray source location
is also much too large to be acceptable. Therefore the x-ray source is most
likely
associated with the H$\alpha$ emission object x-ray 8e, although we have
no spectrum for x-ray 8d which lies marginally closer to the x-ray position.
The spectrum of V410 x-ray 8e indicates that this star is heavily
veiled with a spectral type near M4 or possibly later. Due to the veiling
and the approximate spectral type, the reddening is poorly determined. The
H$\alpha$ emission is very broad. The spectra of all of the other x-ray
candidate objects exhibit H$\alpha$ in emission and M type spectra.

Among the anonymous stars for which we have obtained spectra, there are 4 M
type
stars. One of these stars, V410 anon 13, clearly shows significant extinction,
indicating that it cannot be a foreground star.
This star has an H$\alpha$ emission equivalent width of 22{\AA} indicating
that it is most likely a member of the cloud population of PMS stars.
The other 3 M stars show little or no extinction ($A_V \leq 0.10$) and little,
if any,
H$\alpha$ emission ($W(H\alpha) \leq 2\AA$); therefore they are likely to be
foreground M dwarfs.
The fifth star is an A2 star, showing no H$\alpha$
emission, and a large extinction, 10.45 mag. at V. The strength of the
\ion{O}{1} triplet at 7774\AA \/ is consistent with the spectral type found
from the H$\alpha$ line profile and a luminosity class of V-III.
The H$\alpha$ profile indicates a main sequence star.
At a galactic latitude of $-$15.5 and a galactic longitude of 168.7, only
0.1 A III stars /$\sq\deg$ are expected within the volume formed by the
cloud size and the distance to the star were it a giant. This star is thus most
likely a
main sequence A star at or close to the cloud distance.

We can now examine the distribution of spectral types for the previously
identified cloud population and compare it with that found for the newly
identified population. This is shown in Figure 5. Clearly the {\em newly
discovered cloud members include a much larger number of later type
stars}, indicating that previous studies greatly undersampled
the lowest mass PMS stars.
We can also examine the distribution of reddening values
within the 2 groups as shown in Figure 6. The reddening values in the
previously identified population are heavily weighted toward low values
with a median at 1.25 mag. The median value for the newly identified
population is 3.80 mag. indicating that a major reason for the absence of
these stars from previous surveys is the relatively large extinction
associated with these intrinsically faint objects.

\pagebreak

\subsection{The Infrared Color-Color Diagram}

In Figure 7 we present the near infrared color-color diagram for the
stars in Table 3. The solid lines show the locus of the main sequence
and of the giant branch. The dashed lines show the bounds of the region
within which reddened main sequence stars may fall. Stars lying to
the right of this region show definite evidence for excess emission believed
to arise from circumstellar
accretion disks. Stars lying between these lines may still possess
accretion disks. However it is impossible to distinguish such stars from
reddened main sequence stars or reddened WTTS in this diagram. Also shown
on the lower and left axes of this figure are
the marginal distributions of all of the data for these sources. This
allows us to also indicate the locations for objects too red to have
measurements at all 3 wavelengths and allows quick evalution of the
distribution of points in each color (Tufte 1983).

We may also obtain a measure of the extinction suffered by each star from this
diagram. This is useful, both as a consistency check with the
optical and x-ray data, and to provide extinction estimates for objects
for which we have no other extinction measure. Weak line T Tauri stars
successfully deredden back to the main sequence in this diagram, using
the extinction derived from the optical region photometry. However CTTS define
a
different sequence when the reddening is derived from their optical
colors in the red region and high signal to noise spectra are used to
assign spectral types (Strom, Strom
\& Merrill 1993). Those stars which we are able
to clearly assign to one of these 2 classes we deredden to the
appropriate sequence. We derive
extinctions for the remaining stars, assuming that
all of these stars are WTTSs. This procedure overestimates the extinctions for
those
stars which are actually CTTS owing to the fact that such stars have
intrinsic excess emission arising in circumstellar accretion disks.
For most stars, the extinctions determined by each of these
methods are in reasonable agreement.
Extinctions derived in this way are used only for the 4 objects in Table 3
lacking optical photometry and/or spectral types. The derived values are given
in Table 4.

We can see from this diagram that most of the candidates
for identification with the x-ray sources are clearly PMS stars
exhibiting the excess infrared emission characteristic of circumstellar
accretion disks. The three
apparently most heavily reddened objects are two of the x-ray discovered
sources
and IRAS PSC04154+2823. The only candidate for identification with V410
x-ray 8 which clearly exhibits disk emission is x-ray 8e; it is the object
which also has
the smallest positional discrepancy and which exhibits broad H$\alpha \;$
emission.
The infrared colors of V410 anon 13 also are in agreement with assignment
to cloud membership based on its observed infrared excess. The other reddened
serendipitous source, V410 anon 9, is also a possible cloud member. This
star has a spectral type of A2 and an inferred extinction of 10.45
magnitudes at V. Kim (1990) detected this star in his
near infrared survey and placed it in his definite PMS group from its
location in the infrared color-color diagram. While our colors
are little different from his ($\Delta (J-H) = -0.16$; $\Delta (H-K) = 0.03$),
we feel that this star can
deredden naturally to the appropriate colors for its spectral type,
given the errors in the photometry, although excess infrared emission
cannot be ruled out. There is no reflection nebula
associated with this star and there is no emission at H$\alpha $.
Therefore it is not certain whether this star is associated with
L1495E.

\subsection{The HR Diagram}

One of the major goals of the study of a young star formation region is to
place the stars in the HR Diagram to estimate the masses of the stars and
their approximate ages in order to determine the rate of star formation
versus time in the region. With mass estimates for the stars,
we can calculate the star formation efficiency of the
molecular cloud and compare the local IMF to that observed in the field.
The data in Table 3 enable us to locate the population of L1495E
in the HR Diagram.

{}From our knowledge of the components of the spectral energy
distributions of T Tauri stars (Bertout, Basri, and Bouvier 1988, Hartigan et
al.\
1992), we know that the maximum contribution of the
stellar photosphere to the spectral energy distribution is made at wavelengths
near $1\mu$m.
Flux variations due to rotational modulation are also small at these
wavelengths (Vrba et al.\ 1986, Bouvier \& Bertout 1989).
Therefore we will choose to use the flux at I or J as
our measure of the stellar photospheric flux. When data at J are
available, J is preferred. We must still estimate the
amount of extinction for each star, although we have minimized the importance
of
extinction corrections by working this far to the red. The spectral type and
R$-$I
color yield a color excess for each star, E(R$-$I)~=~1.4~$A_J$,
which, when combined with a standard reddening law, becomes an
extinction at the chosen wavelength, I or J ($A_J$~=~0.46~$\times
A_I$~=~0.28~$\times A_V$).
Using the bolometric corrections
of Bessell \& Brett (1988), we find that the bolometric correction at I
varies rapidly as a function of spectral type beyond M0, while the
variation at J is slow and monotonic over the entire spectral range of
interest. Therefore we will use the J
magnitude, whenever possible, (corrected for extinction, distance (m-M = 5.88)
and bolometric correction) as our measure of the stellar luminosity.
Using the temperature-spectral type calibration of Hartigan et al.\ (1994)
compiled from the work of Bessel \& Brett (1988) and Schmidt-Kaler (1982),
we derive an effective temperature for each star for which we have a
spectrum. The errors in placement of stars in the HR Diagram vary with
spectral type. They can be estimated by using the values given below for
two spectral types.
  \[ K5 \pm 1~\mbox{subtype} \left\{ \begin{array}{rlrlr}
     +0.019  & \mbox{in $\log T_{eff}$} & -0.03 & \mbox{in $R_c - I_c$ \& } &
Observed \\
     -0.020  &    & +0.11   & \mbox{E($R_c - I_c$)} \\
	     &    &         &           \\
     -0.02   & \mbox{in $A_J$} & -0.05 & \mbox{in $BC_J$} & Derived  \\
     +0.08   &    & +0.01   &            \\
	     &    &         &             \\
     +0.01   &  \mbox{in $\log L$ }  & &  & Result \\
     +0.03   &         &   &
     \end{array}
     \right. \]

  \[ M3 \pm 1~\mbox{subtype} \left\{ \begin{array}{rlrlr}
     +0.019  & \mbox{in $\log T_{eff}$} & -0.21 & \mbox{in $R_c - I_c$ \&  } &
Observed \\
     -0.027  &     & +0.26   & \mbox{E($R_c - I_c$)} \\
	     &     &         &           \\
     -0.15   & \mbox{in $A_J$} & -0.08 & \mbox{in $BC_J$} & Derived \\
     +0.19   &     & +0.08   &           \\
	     &     &         &            \\
     +0.03   & \mbox{in $\log L$ }  &  & & Result \\
     -0.04   &      &  &
     \end{array}
     \right. \]

However, the sometimes large light variations of PMS stars prove to be the
largest source of random error in the placement of these stars in the HR
diagram (Hartigan et al. 1994). This can be seen immediately in Table 3
where where our 2 observations of V410 x-ray 1 differ by more than a magnitude
at J and of V410 x-ray 3  differ by more than 0.5 magnitude.
There are multiple sources for the variability
exhibited by PMS stars. Rotational periods have been determined for many
PMS stars by observation of the spot modulation of their light curves. The
amplitude of this variation may be as small as a few hundreths of a magnitude
or as large as 1 magnitude (at V) (AA Tau, Vrba et al. 1988). These stars are
also known to have an irregular component to their variability with
very large amplitudes sometimes exceeding several magnitudes
(Herbig and Bell 1988). These irregular fluctuations are presumed to be due to
variations
in the accretion rate through the disks of these stars (Kenyon et al. 1994).
In the extreme case, an FU Ori event may cause a rise of more than 5 magnitudes
over a relatively short period of time (weeks) with a much longer decay time
(years) (Hartmann, Kenyon and Hartigan 1993). Hartigan et al. (1994) found that
the error in placment of PMS stars in the HR diagram was dominated by the
variability of these stars. They estimated that variability was responsible
for $\sigma \simeq 0.05$ in $\log L$ although there is a long tail at higher
amplitudes to the distribution. When all of the possible sources of error
are combined in quadature, the result is a ``typical''
$1\sigma$ uncertainty of 0.08 in $\log L$.

The data placing each of our stars in the HR Diagram is given in
Table 4 and they are plotted in Figure 5 with the $10^8$ yr isochrone
from the D'Antona \& Mazzitelli(1993) evolution tracks. The objects V410 x-ray
2, 4,
PSC04154+2823 and PSC04158+2805 are
placed in the HR Diagram using the (uncertain)
extinctions estimated from their near infrared colors, a bolometric
correction and effective temperature assuming they are middle M stars
similar to the other objects found in this sample.
For V410 anon 3 \& 17, for which there was no near infrared photometry, we
used I as the standard wavelength. The final luminosities for the 5
objects known to be binaries are then
corrected for the light contributed by any secondary star discovered by
high spatial resolution observations (see section 5.3).
These final values are given in the column L(J$_c$).

It is obvious from Figure 8 that the stellar population of L1495E is very
young. However, it is difficult to evaluate exactly how young and how coeval
this population is because different sets of ``modern'' evolutionary
tracks (D'Antona \& Mazzitelli 1993, Swenson et al.\ 1993) show
sufficient differences to make interpretation difficult.  Also, the
isochrones are in intervals of $\log$ Age; therefore the stars are evolving
very rapidly in the upper part of the HR diagram and more slowly as they
age. In Figure
9 we show the data for these stars plotted on a) the D'Antona \&
Mazzitelli tracks using the Alexander opacities and the Canuto and
Mazzitelli convection formulation; and b) the Swenson et al.\ tracks. It
is clear that, for stars this young ($\leq 10^6$ yrs) the tracks differ
sufficiently to influence our estimate of coevality as well as the
"absolute" age. On the D'Antona \& Mazzitelli tracks, for stars with
masses less than 0.5 $M_\odot$, the population appears approximately coeval
with an age of $t \sim 5 \times 10^5\,$yrs. Using a program written by
Patrick Hartigan which
interpolates between the isochrones to assign an age to each star,
we find that 20 of the 22 low mass PMS
stars, excluding those viewed by scattered light (see later in this section,
and the anonymous stars,
have ages $\leq 7.1 \times 10^5$ years.
On the Swenson et al.\
tracks, for the same mass range, the apparent age increases as the mass
decreases and the mean age for the L1495E population is $10^6\,$yrs.
Given the uncertainties in the
model isochrones and the fact that the models are still sensitive to initial
conditions at these early ages, this group can be considered coeval to
within $\leq 1$Myr. The two apparently older stars merit discussion.
We have only one complete set of colors for FQ Tau. However, at the time of
the Leinert et al. (1992) speckle observations, FQ Tau was more than 0.5 mag.
brighter at K than when we observed the star. If this brightness change is also
reflected at J, this star
is then $< 1\sigma$ above the mean age for the group.
The remaining star lies 3.75$\sigma$ above the mean age of the group; this is
not unreasonable for a sample this size.

One possible problem apparent from both sets of tracks is the
gradually increasing age found for stars with masses $\geq 1.5 M_{\odot}$.
While RY Tau and HDE 283572 lie off the face of the cloud, V892
Tau (Elias 1) is heavily embedded in the cloud core and illuminates a
reflection nebula. It is difficult to believe, given the apparent
velocity dispersion of L1495E of $\sim 3.0$ km/s, or 3.0 pc$/10^6$
yr,that this star is over $3 \times 10^6$ yrs old as the tracks imply. A star
formed in the
center of the cloud would have moved more than $200\arcmin$ away from
the cloud center in $3 \times 10^6$ yrs. The motion of this star would
have to be aligned within 10\deg$\,$ of the line of sight and be moving away
from us in order to remain located in cloud center and still exhibit an
extinction of 9 mag. The only alternative (other than suspecting a systematic
error in the tracks) is that the age of this star is correct and that
star formation actually began $3 \times 10 ^6\,$yr ago. In this case the
timescale for
"building"  a star via infall and accretion, before it appears on these tracks,
would have to be much longer for low mass
stars.

{}From examination of Figures 8 and 9 and Table 4, several objects
deserve special mention. There are two members of the previously identified
PMS population that fall well below the main body of the distribution.
The images of
these objects, CK Tau 1 and PSC 04158+2805,
appear definitely extended  on our deep R and I frames, the
PSC object appearing triangular and the object near Hubble 4 appearing
flamelike (Figure 10). We suspect that we are not seeing these
objects directly but instead via light scattered through a circumstellar
envelope or from nearby molecular cloud material. Recent submillimeter
observations of low
luminosity embedded sources in the Taurus region (Barsony \& Kenyon
1992) indicate very large extinctions for these sources, (50 $< A_V <$
600 ), sufficient to
render them undetectable even at $2\mu m$. However these objects
are nevertheless visible in the near infrared and sometimes at even
shorter wavelengths. This paradox can be resolved if the
density distribution surrounding these objects is highly asymmetric.
Models of sources such as these using asymmetric density distributions have
been
developed recently by Kenyon, Calvet \& Hartmann (1993) and Kenyon et al.\
(1993). These
authors find it necessary to introduce a bipolar hole in the density
distribution in order to reproduce both the observed spectral energy
distributions and the extended images of such sources.
Both the spectral energy distributions and direct images of the objects
shown in Figure 10 are very similar to those represented by these models.
The (R$-$I) colors of
these objects indicate 5-10 magnitudes of extinction.
However, the models of Kenyon et al.\ (1993), while only calculated for the
J,H, \& K bands, suggest that the actual extinction that
should be applied to the observed magnitudes could be a factor of
3 or more higher than that inferred from the observed near infrared colors.
The discrepancy should be even greater for extinctions determined from
the R-I color.
The visible and near infrared light is
due to scattering and not transmission, and therefore the anomalous
location of these stars in the HR diagram can easily be accounted for by an
underestimate of the extinction.

Our spectral type for CK Tau 1 is uncertain as the spectrum appears
heavily veiled
and contains strong emission lines of H, [\ion{O}{1} ], [\ion{S}{2} ],
[\ion{Fe}{2} ] among others. This contributes a major source of error in the
reddening estimate as we are probably not measuring photospheric colors.
Our previously published images of CK Tau 1 in H$\alpha $ (Strom et al.\ 1986)
indicated a faint bipolar jet emanating from this star.
The recent narrow band imaging by Goodrich (1993) confirms this result,
showing a bipolar jet clearly in the [S~II] images.

Among the x-ray discovered sources in this cloud, it is clear that V410
x-ray 1, 2, 3, 4, 5a, 6, 7, and 8e are member of the PMS population of L1495E.
In most cases they are the only possible optical object which could be
identified with the x-ray source. The near infrared colors of these
objects indicate excess infrared emission. When we have spectra for these
objects, H$\alpha $ appears in emission ($W(H\alpha ) \geq 8\AA $).
As discussed previously, x-ray 8a is a background M giant seen through
the edge of the cloud. We can make no statement about the possible
association of V410 x-ray 8b, 8c, 8d, 8f and 5b since we have no spectra
for those objects.

Of the serendipitously observed stars, V410 anon 3, 12 \& 17 show little or
no extinction or H$\alpha $ emission. Their derived luminosities (at the
distance of the Taurus clouds) place them above the main sequence.
However they are more likely {\em main}
sequence stars in front of the cloud.  The galactic model (Bahcall and
Soneria 1980) predicts $\sim$ 10 M stars between us and the Taurus
clouds projected on the central region of the cloud ($r \leq 25
\arcmin$). The other two serendipitously
observed stars may well be members of the cloud population. V410 anon 13
has an H$\alpha $ emission equivalent width of $22\AA \:$ and an inferred
extinction of 2.75 magnitudes at V. This star is clearly a strong
candidate for inclusion in this young group. The lack of detection in
the ROSAT image is probably due to its low luminosity, below that of any
of the x-ray detected objects. The star V410 anon~9, discussed
previously, appears to be a normal A2 star with no H$\alpha$ emission.
When dereddened and placed at the distance of the Taurus clouds, the
star lies just above the main sequence, not far below V892 Tau.
It is possible that this star lies behind the cloud, but it cannot lie far
behind because the probability of
finding an A giant this far above the galactic plane is small.

\subsection{Binaries in L1495E}

While we have approximately doubled the previously known population
within the central region of L1495E, all with M stars, it is still possible
that even more low mass stars may be hidden as secondaries in binary systems.
Fortunately there has been much recent high spatial resolution work
which allows us to evaluate this possibility (Ghez 1993, Simon et al.\
1992, Leinert et al.\ 1992, Simon 1992, Ghez, Neugebauer \& Matthews
1992).

Of the previously identified PMS sample (19 stars), 7 are known to be
binaries (V410 Tau, DD Tau, CZ Tau, V892 Tau, FO Tau, FQ Tau, and LkCa
7). Eight stars were found to be single at the separations, flux ratios (and
sometimes position angles) to which the measurements were sensitive. Four
stars (CK Tau 1, PSC04154+2823, PSC04158+2805 and LkCa 5) were not examined.
V892 Tau was discovered to be a binary by
Thornley et al.\ (1989) using near infrared imaging, The secondary was later
confirmed
by Skinner et al.\ (1993) in their VLA observations of this region. They
detected both V892 Tau and its close companion to the NE as well as
CK Tau 1 and Hubble 4 (Skinner 1993).

While a few B and A stars in young stellar groups have been identified as
possible x-ray sources (Caillault \& Zoonematkermani 1989,
Strom et al. 1990, Pravdo \& Angelini 1993), there is always the possibility
that the x-ray emission is attributible to a low mass companion (Schmitt et al.
1985).
In this case, it is reasonable to attribute the x-ray luminosity of V892 Tau
to the low mass secondary of the system. The derived x-ray luminosity
is consistent with the inferred luminosity of the secondary if we assume that
the luminosity ratio of the two components of V892 Tau is close to the ratio
of their infrared luminosities, the star lies on the mean isochrone for
this group, and we use the bolometric correction inferred from this placement.

Flux ratios are in general available at only one wavelength for these
binaries, usually K (2.2$\mu$m). For 5 of the 7 known binaries, the flux
ratios at K lie between 1.1 and 2.2. The other 2 stars (V892 Tau and
V410 Tau) have relatively much fainter companions.
If we assume that this ratio represents the true
luminosity ratio of the components, the extinction for both components
is the same, and that they lie on the $10^6$ yr isochrone, we can estimate
the masses of these stars. Therefore these 7 companions can
also be added to the low mass ($M < 0.3 M_{\odot}$) population of the
cloud.

\subsection{The Cloud IMF}

Among the basic problems to be addressed by studies of star formation are
(1) to derive the initial mass function (IMF) for isolated star formation
regions and compare with that derived from study of stars in the solar
neighborhood; and (2) the local efficiency of star formation in the star
forming cores. Implicit in the determination of the IMF is the question
of whether stars of mass lower than that in which hydrogen burning can
be sustained (brown dwarfs) are plentiful and may therefore contribute
to solving the ``missing mass'' problem.

It is now possible to estimate the star formation efficiency of the
L1495E cloud. The mass of the cloud has been estimated at between 180
and 230 $M_\odot$ (Ohnishi \& Mizuno 1993). The total stellar mass in
the central region of the cloud depends slightly on the
set of tracks used to make the mass estimate. If the Swenson et al.\
(1993) tracks are used, the total stellar mass is estimated at $\sim 13
M_\odot$; if the D'Antona \& Mazzitelli (1993) tracks are used, the
total stellar mass is estimated at $\sim 10.5 M_\odot$. This leads to
star formation efficiencies between 5\% and 7\%. If the stars in the
outer region of the cloud are included, the star formation efficiencies
increase to between 8\% and 9.3\%.
Current constraints on the formation of bound clusters (Lada, Margulis \&
Dearborn 1984) require that at least 30\% of the cloud mass must be converted
to stars. Unless a large number of low mass members of this group remain
hidden within the cloud, at luminosities below the limit reached by this ROSAT
exposure, it is unlikely that this group will remain bound.

One of the first studies to examine the distribution of spectral types in
nearby star forming regions was carried out by Cohen \& Kuhi (1979). While a
small
number of stars later than spectral type M0 were found in Taurus, those authors
assumed that the flattening of the mass spectrum at low masses apparent
in their data, was a selection effect due to a magnitude cutoff. Since
that work, there have been several surveys aimed at increasing the known PMS
population
of the Taurus clouds: the proper motion surveys of Jones \& Herbig
(1979), Hartmann et al.\ (1991), and Gomez et al.\ (1992); the \ion{Ca}{2} H,K
emission survey of Herbig et al.\ (1986); the x-ray sources from {\it
Einstein} pointed observations (Walter et al.\ 1988); and the deep
H$\alpha$ emission survey of Brice\~{n}o et al.\ (1993).
The recent proper motion surveys, (Hartmann et al.\ 1991, Gomez et al.\
1992) have found a small number (7) of additional members of the Taurus
PMS population, usually late type stars (SpT later than M3) with relatively
weak
H$\alpha$ emission. Most recently, Brice\~{n}o et al.\ (1993) have used very
deep
(V $\approx 18$) objective prism H$\alpha$ emission surveys to search
for as yet undetected pre-main sequence stars in the Taurus clouds.
They identified 12 new T Tauri stars, 5 of which were found in L1544,
a region previously unsurveyed in H$\alpha$ and containing
only one identified PMS star. Most of the newly discovered
PMS stars are classical or strong emission T~Tauri stars (CTTS).
The PMS
evolutionary tracks used until this point (Cohen \& Kuhi 1979, Strom et
al.\ 1989b) have identified the peak of the N(SpT) distribution in the
Taurus clouds (at K7-M0) with stars
of $\sim 1 M_\odot$ with the number of stars decreasing with mass until
few stars are found below 0.4 $M_\odot$.  This result is in marked
disagreement with the IMF derived for the solar neighborhood.
The IMF found for the local solar
neighborhood (Scalo 1986, Kroupa, Tout \& Gilmore 1990, Tinney, Mould \&
Reid 1992) can be well fit by a log normal function peaking at $\log M =
-0.5$ and then either decreasing as the mass decreases or possibly staying
constant for all lower masses.
The PMS stars identified in the above surveys have spectral types with either M
or continuum
+ emission characteristics.
Their R magnitudes range from 14.2 to 18.5.
Even though most of these newly discovered stars are M stars,
the apparent deficiency of low mass stars remained.

One possibility for hiding additional low mass stars is as secondaries in
binary and multiple systems. Recent studies (Ghez 1993, Simon et al.\ 1992,
Leinert et
al.\ 1992) have suggested that most stars form in multiple
systems. This possibility must also be evaluated in order to fully account
for the IMF in star forming regions.

It is also important to note that the masses for these stars
obtained from the ``modern'' tracks are quite different from those
obtained in the past. Here stars that were recently considered as
typical 1 $M_{\odot}$ PMS stars are now found to be 0.3-0.5$M_{\odot}$
stars. While this may seem a small change, these typical TTS will not be
G stars when they arrive on the main sequence but early M stars. This
change may have a significant effect on the analyses of disk stability and
binary period
distributions (Ghez 1993, Leinert et al.\ 1992) since derived ratios of
disk mass to stellar
mass will increase and the total system masses for binaries will decrease.

As a consequence of (1) the newly discovered low mass members of the L1495E
cloud population (x-ray detections, 8 stars; anonymous stars, 1 star;
low mass companions, 7 known), and (2) masses derived from new tracks,
the apparent deficit in the population of the low mass end of the
IMF in this part of the Taurus clouds will be decreased considerably
(see below).

The combination of newly observed and already known PMS stars in L1495E
enables a modern discussion of the IMF. Since we have a deep
survey only in the central region of the ROSAT field, we will use only
the objects (known and newly discovered) that fall within that area. We begin
with the 12
previously known members of the cloud population. To this we add
8 x-ray discovered PMS stars, 1 certain PMS star from the group of
anonymous optically visible stars projected upon the cloud, 1 possible
addition to the intermediate mass star population associated with the
cloud, and, finally, the 5 known secondary stars to the cloud members
within the central region
that have been observed at high spatial resolution. This gives a total
of 27 probable cloud members within a circle of radius 25\arcmin$,$
centered on V410 Tau.

In order to construct an IMF for the PMS population of this cloud, we
must infer a mass for each star from its position in the HR Diagram
relative to the calculated evolution tracks. Here we must deal with
the problem that the mass deduced for each star differs depending on
the tracks we choose to use. This is particularly true at the
very low mass end of the range in which we are interested, where the two
sets of tracks are most divergent. We construct an IMF for each set
of tracks in order to better compare the derived shape of the cloud
IMF. The results are presented in figure 11. The results from both sets
of tracks yield IMFs that peak at $\log M = -$0.5, identical to the
result found by Scalo(1986) from solar neighborhood stars.
There are too few stars of masses greater than $\sim1 M_{\odot}$
to allow us to draw any conclusions about the higher mass end of the IMF in
L1495E, and the behavior of the derived IMF for masses below the peak is
dependent upon the tracks chosen. While the IMF drops toward
lower masses, the drop is much more severe for the D'Antona
\& Mazzitelli tracks. If the Swenson et al.\
tracks are used, 10 stars are found in the 2 bins of
lowest mass ($ -1.2 \leq \log M \leq$ -0.8), while if the D'Antona \&
Mazzitelli tracks are chosen, only 2 stars
are found in these bins. For this reason, and because (1) less than half of
the cloud members have been searched for companions of subarcsecond
separation; and (2) there is no knowledge of binary systems with
separation below the sensitivity of current instrumentation, we are
unable to make a definitive statement about the shape of the cloud IMF
at these low masses. However, there is {\em no evidence for a rise toward
lower masses} if the PMS population we have identified is either complete
or representative.

We have attempted to estimate the number of possible additional members of the
cloud population
from our near infrared images. In addition to deriving the JHK
magnitudes for the candidate x-ray sources, we also obtained near
infrared magnitudes for all other objects appearing within these frames.
Most of these objects, in the frames away from the cloud edge, are quite
faint. Consequently, the photometric errors are larger and their colors are
less definitive than we would
like for the purpose of identifying possible cloud members. However, there
are clearly some stars (8) whose
presence in the central area of the cloud and whose colors (indicating
the presence of disk emission) suggest that they are members of
the cloud population. One of these stars, not visible on our long I
image, just 40\arcsec$\;$ south-southeast of V410 Tau, is brighter than
$K = 10$ and extremely red. Some of the stars in the region near V892 Tau
are so red that no J magnitudes can be obtained, however they cluster
about the tight group of stars found in this region, suggesting membership. A
table
giving the approximate coordinates of these stars and their photometry
can be found in the Appendix.

We must emphasize again the need for very deep x-ray searches of star
formation regions in order to locate the low mass members of the PMS
population. The lack of other comparably deep observations of star
forming regions with ROSAT while the PSPC was operative is extremely
unfortunate.

Other nearby star formation regions have not shown a deficiency of
M stars comparable to that seen in Taurus-Auriga.
Hughes and Hartigan (1992) compare the spectral type distribution in the
Chamaeleon II cloud with that seen in Taurus. They find a higher fraction
of stars with late spectral types in Cha II than are seen in Taurus.
Krautter \& Kelemen (1987) show the distribution of spectral types for
the Lupus star formation region, at a comparable distance, which has only 2
stars earlier than K5
while 47 stars later than M0 are found.
The new PMS evolution tracks can explain the difference in the spectral type
distribution between the Taurus and the Lupus clouds. The mean age of
the stars in the Lupus clouds is considerably older than that of the stars
in the Taurus clouds (Hughes \& Hartigan, private communication).
Therefore, using the ``modern'' tracks, as shown in Figure 9,
the mean spectral type of the
population will be considerably later in Lupus than in Taurus for the
same distribution of stellar masses since these tracks take a
considerable turn to lower effective temperature after $\sim 1 \times
10^6$ yrs.

\pagebreak

\section{Correlations of X-ray Properties with the Stellar Parameters}

Because the x-ray emission from PMS stars appears to be consistent with
coronal emission models, the strength of the x-ray emission should be
correlated with the strength of the dynamo generating the stellar
magnetic field. In turn, we expect that a measure of the rotation of the
star will be the proper surrogate for the dynamo strength; and therefore
that stellar rotation should correlate with the x-ray emission.
However, one
complication in this chain of logic is the fact that these very young
stars are completely convective, and the currently favored dynamo models
operate at the interface between the convective and the radiative zone,
driven by differential rotation. However there is a distributed dynamo
model (Rosner 1980) that works throughout the convection zone.

In this section we will evaluate the simplest correlations that we can
make, the correlation of the x-ray luminosity with the stellar
luminosity and with the rotational velocity. In these correlations, we
will omit the data for V892 Tau since it is likely that the x-ray
emission is due to the fainter component of the system for which we are
unable to establish a reliable luminosity estimate since we have no spectral
type.

Our x-ray data consist solely of measured values with no upper limits,
therefore we have no need to use survival statistics methods in our
analysis. The x-ray luminosity and
the stellar luminosity are very strongly correlated, with a formal probability
of $<$0.001 that they are uncorrelated. The slope of the relationship
between $\log L_x \;$ and $\log L_\star \;$ found from a
least squares solution is $1.80 \pm 0.01$. However, since the values of
$\log L_x \;$ are instantaneous snapshots of highly variable sources,
the accuracy of the slope determination is  no doubt worse than that
quoted. This relationship
is shown in Figure 12. The WTTS, CTTS and stars that
are no longer fully convective are indicated. While there is considerable
overlap between the CTTS and WTTS, the WTTS are clearly systematically
more luminous than the CTTS. While it is true that some of the WTTS are
found at earlier spectral types than are the CTTS and therefore at
higher stellar luminosities, it also appears to be true that the WTTS
are more luminous at x-ray wavelengths than the CTTS when compared at
constant stellar luminosity. DD Tau
is plotted twice in this figure, at its pre-flare and flare locations.
The pre-flare location falls just below the center of the band, in the
midst of a group of other points, while the flare point lies well above
the majority of the points. It is possible that the other points lying well
above the major part of the distribution represent stars that were not
in their quiescent state at the time of the observation.
It is also interesting to note that the x-ray luminosities of the least
luminous of these PMS stars have x-ray luminosities comparable to those
of the young near main sequence K and M stars of the Pleiades (Stauffer et
al.\ 1993). In fact, because our detection limit is lower than
that of the Pleiades observations, some are even less luminous than the
upper limits found for the undetected Pleiades members. This must be
taken into consideration when testing dynamo models and angular momentum
evolution scenarios.

We can compare this distribution with that found for the PMS members of
the Cha I cloud (Feigelson et al.\ 1993). These authors find a
correlation with a slope of $1.00 \pm 0.15$ for the relationship between
the stellar luminosity and the x-ray luminosity for
the Cha I stars. For the stellar luminosities they used the values found
by Gauvin \& Strom (1992) adjusted for the closer distance that they
chose to use. Since in this relationship the distance enters in the same
manner on both axes, it does not matter which distance is chosen. However,
the method by which the stellar luminosity is calculated has now been
improved. In order to remove any systematic effects introduced in this
way, we have recalculated the stellar luminosities for the Cha I stars
using the J magnitude as our measure of the stellar photosphere as discussed
in section 6.2 and
using the same data compiled in Gauvin and Strom (1992). The x-ray
luminosities are taken from Feigelson et al.\ (1993). In Figure 13 we have
plotted the data for Cha I with the relationship derived for
the stars in L1495E and
shown in Figure 12. We can clearly see that the Cha I CTTS, fainter WTTS
and radiative track stars
exhibit the same slope to this relationship as found for the stars in L1495E,
however they have systematically lower x-ray luminosities at constant
stellar luminosity. The
stars deviating from this relationship are the most luminous stars in the
cloud (excluding the 2 A stars not considered here). Most of these stars
are no
longer fully convective; the most luminous WTTS lies on
the boundary of the fully convective region. The only real anomalies are
the 2 CTTS (VW Cha and WW Cha). One of these, VW Cha, is a binary with a
separation of 17\arcsec $\;$(Reipurth and Zinnecker 1993, Schwartz 1977),
both components of which must lie within the x-ray beam. Reexamination
of the spectra of these stars (Appenzeller, Jankovics \& Krautter 1983)
shows that these stars would better be classified as continuum + emission
stars leading us to overestimate the stellar luminosity for 2 reasons,
1) the flux at J is contaminated by disk emission, and 2) the assigned
early K spectral types are most likely too early leading us to use a
bolometric correction too small. There is no simple way to make a
correction to the luminosities assigned to these stars, short of
obtaining high resolution spectra. Therefore we merely indicate that the
stellar luminosities for these stars are upper limits. The relative
absence of WTTS in this region is no doubt due to the fact that the
stars with adequate photometric and spectroscopic data are those which
were discovered by H$\alpha$ emission surveys. Many more cloud members
have been discovered in the x-ray emission surveys and in more sensitive
H$\alpha$ emission surveys (Hartigan 1993), but as yet lack the
information necessary to place them in the HR Diagram.

An interesting aspect of this diagram is that the x-ray luminosity of
stars which are now on their radiative tracks appears to be decreasing faster
with the age of the star than do the x-ray luminosities of stars still
fully convective. This could reflect the fact that the surface magnetic
field strength is relatively weaker for these stars and therefore that
the dynamo generation mechanism is not as efficient as that in the fully
convective stars. A possible correlated effect has been seen by Edwards
et al. (1993) in the angular momentum statistics of these stars. While
the circumstellar disk appears able to regulate the rotation period of
the fully convective stars quite efficiently through the linking of the
stellar magnetic field with the disk, this process appears to be less
efficient in stars that have radiative cores.

Another way to look at the correlation of rotation with coronal heating
is to use the relationship first proposed by Bouvier (1990) between
x-ray surface flux and rotational period of the star. We are
fortunate to have 6 rotational periods determined for the stars in
L1495E (Rydgren et al.\ 1984, Bouvier et al.\ 1986, Vrba et al.\ 1986,
Walter et al.\ 1987, Vrba et al.\ 1988, Bouvier et al.\ 1993), and
also v$\sin i$ data (Hartmann et al.\ 1986, Hartmann \& Stauffer 1989)
for this group. There are many
fewer rotational periods determined in Cha~I (Bouvier \& Bertout 1988).
In order to make use of
a larger database, we will translate the rotation periods into
projected
equatorial velocities through knowledge of the stellar radius. In Figure
14a we show the relationship between $\log F_x$ and rotational velocity.
Those velocities that are derived from rotation periods are shown by
filled circles while those known only from measurement of stellar line
widths are marked with open circles. In Figure 14b the same relationship
is shown for the Cha I cloud stars; the v$\sin i$ values are taken from
Franchini et al.\ (1988) and Walter (1992). The two stars marked with limiting
values in both rotational velocity and surface flux are VS Cha and WW
Cha, mentioned earlier. These stars are very heavily veiled and therefore
the luminosity and thus the radii are probably over estimated, leading
to an underestimate in x-ray surface flux. Franchini et al.\ (1988) note
that the rotational velocities that they have determined would be over
estimated if a substantial veiling flux is present.
In Figure 14a the data for the
Pleiades stars is shown (Stauffer et al.\ 1993). The Pleiades
data can be described as filling a wide band in log $F_x$ at rotational
velocities $>20$ km/s. Below that velocity much lower values of log~$F_x$
can be found. It is difficult to look for a functional relationship in
this data since values of v$\sin i$ give only lower limits to $V_{eq}$.
However, it is clear that the slowly rotating Pleiades stars exhibit much lower
x-ray
surface fluxes (Stauffer et al.\ 1993). The L1495E x-ray data is consistent
with the Pleiades data. However, because of the small number of stars in
this group having measured v$\sin i$'s, only 1 slow rotator, no independent
assessment of this behavior can be made for PMS stars from this data.
Therefore we cannot separate age and rotation effects.

Because the derivation of other stellar parameters, in particular the
stellar mass, are dependent upon the set of stellar evolutionary tracks
used, we prefer to wait until the new tracks have been further evaluated
until attempting further correlations with derived stellar parameters.
It will also be useful to have a larger body of x-ray data derived from
the ROSAT pointed observations for stellar populations of a range of
ages. We will then be able to separate the effects of age and rotation for
stars in a small mass range from first appearance to the main sequence.

\section{Conclusions}

The L1495E cloud is a region of very active star formation. Our multiwavelength
deep survey of this region has considerably increased our knowledge of the
PMS population within the cloud. The ROSAT PSPC observation allowed us
to identify 8 new low mass ($0.08 M_{\odot} \leq M \leq 0.6 M_{\odot}$)
members of the cloud population. Serendipitous observations of otherwise
undistinguished stars has added 1 (or possibly 2) more members.  Nearly
all of the new members of the cloud population are of spectral type M3
and later and thus represent a heretofore unrepresented fraction of the
cloud population. This
population appears to be nearly coeval; that is the spread in the HR
Diagram is small at any given mass. The mean age lies between $5 \times
10^5$ and $1 \times 10^6$ yr depending upon the adopted tracks.

These observations combined with the use of ``modern'' evolutionary
tracks allow us
to address an outstanding problem concerning the PMS population
in the Taurus molecular clouds,
the apparent deficiency of stars with masses below $\sim 0.40
M_{\odot}$.
While the addition of the newly discovered cloud members and the identification
of the low mass companions of other cloud members helps us to
alleviate this deficiency, the use of the new evolutionary tracks is the
key factor in understanding the mass distribution in the L1495E
population. The masses deduced for stars of $1 \times 10^6$ yrs and
younger are considerably lower than those previously derived (Strom et
al.\ 1989b). This allows us to understand the large difference in the
effective temperature distribution, as reflected in spectral types,
between the PMS population in Lupus and that in Taurus as merely a
reflection of different ages of the two groups.

The near infrared observations of the newly recognized cloud members
show that 6 of 8 exhibit colors characteristic of disk
emission. With the spectra obtained with HYDRA, we find
that, with the possible exception of the serendipitously observed A2
star, the new cloud members are low mass stars ($M < 0.6 M_\odot$) and
are typically deeply embedded in the cloud. The images of 2 of the low
luminosity IRAS sources are extended. These images, in combination with
their spectral energy distributions, and the models of Kenyon et
al.\ (1993), demonstrate that these objects are still surrounded by dense
core material, even though one of the objects is actually found at the
edge of the main cloud.

Although there are only 35 known members of this stellar aggregate, 27
of which lie within the completely sampled region, we
show that the observed IMF for this group is consistent with the IMF
found by Scalo (1986) for the field population, with a peak near $\log M
= -$0.5 and a falloff toward lower masses. There is no evidence for a
rise as mass decreases.

The x-ray luminosity of the L1495E stars is highly correlated with
the stellar luminosity with a slope to the relationship of 1.80 $\pm$
0.01. This can be seen in the
observations of the Cha I stars as well, although the higher mass stars
in this group have much lower x-ray luminosities than their counterparts
in L1495E, perhaps due to the greater age of the Cha I association ($\sim 2
\times
10^6$ yr on the Alexander tracks).
When the x-ray surface flux of the Pleiades sample
is examined as a function of the rotational velocity of the stars, it is
seen that no systematic variation is found for velocities greater than 20
km/s. However, below that velocity the range of observed x-ray surface
flux increases dramatically with stars of low projected rotational
velocities showing the lowest surface fluxes. When the rotational
velocity drops below 20 km/s, the energy source powering the x-ray
emission in these stars is substantially diminished. The x-ray data for
the L1495E stars is consistent with the Pleiades data. However there are
too few stars in the sample having measured rotational velocities to enable
us to separate age and rotation effects.

\acknowledgements

The authors would like to thank Patrick Hartigan for obtaining the R and I
band CCD frames, and Ronald Probst, Michael Meyer, Patricia Knezak and Lynne
Hillenbrand for each
obtaining part of the near infrared frames. Without their help this
paper would have greatly suffered.
The authors would also like to acknowledge support from the National
Science Foundation (SES), the NASA Planetary Science and Origins of Solar
Systems programs, the NASA ADP
program and, of course, the ROSAT observations support (KMS).

\section{Appendix}

In this section we give the data on the anonymous stars that were
serendipitously observed with HYDRA and with SQIID. In Table 5 are given
the positions and  photometric data for the optically visible stars
which are seen projected on the face of the cloud. Five of these stars
are discussed in the text.

In table 6 are given approximate positions and near infrared photometric
data for stars appearing in our SQIID frames whose colors indicate that
they might be cloud members.

\begin{figure}

\caption{A finding chart for the x-ray discovered PMS stars in the L1495E
cloud. Also identified are the previously known PMS stars falling within
the CCD frame. This image is 23\farcm5 on a side.}

\end{figure}

\pagebreak

\begin{figure}
\plotfiddle{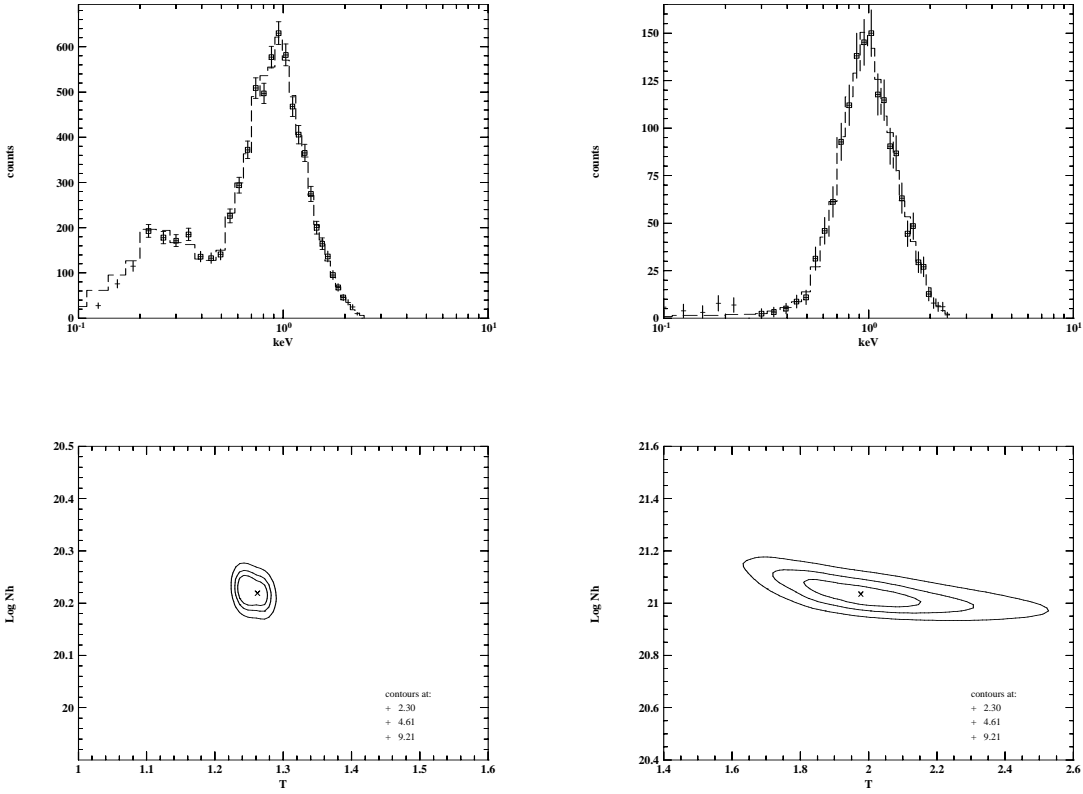}{12.0cm}{-90}{60}{60}{-250}{360}

\caption{Examples of model x-ray spectrum fits to 2 high signal to noise
observations. In the left hand pair of plots is shown the fit to the
x-ray spectrum of V410 Tau, a very lightly reddened object on the front
surface of the cloud. In this case both the temperature (in KeV) and the column
of neutral hydrogen are well constrained as seen from the $\chi^2$ plot
on the lower left. In the other case (Hubble 4), the softer x-rays have
been absorbed by the intervening cloud material. The lack of
information at these energies leaves us with a fit that constrains the
amount of intervening material but imposes less constraint on the
temperature.}

\end{figure}
\begin{figure}
\plotfiddle{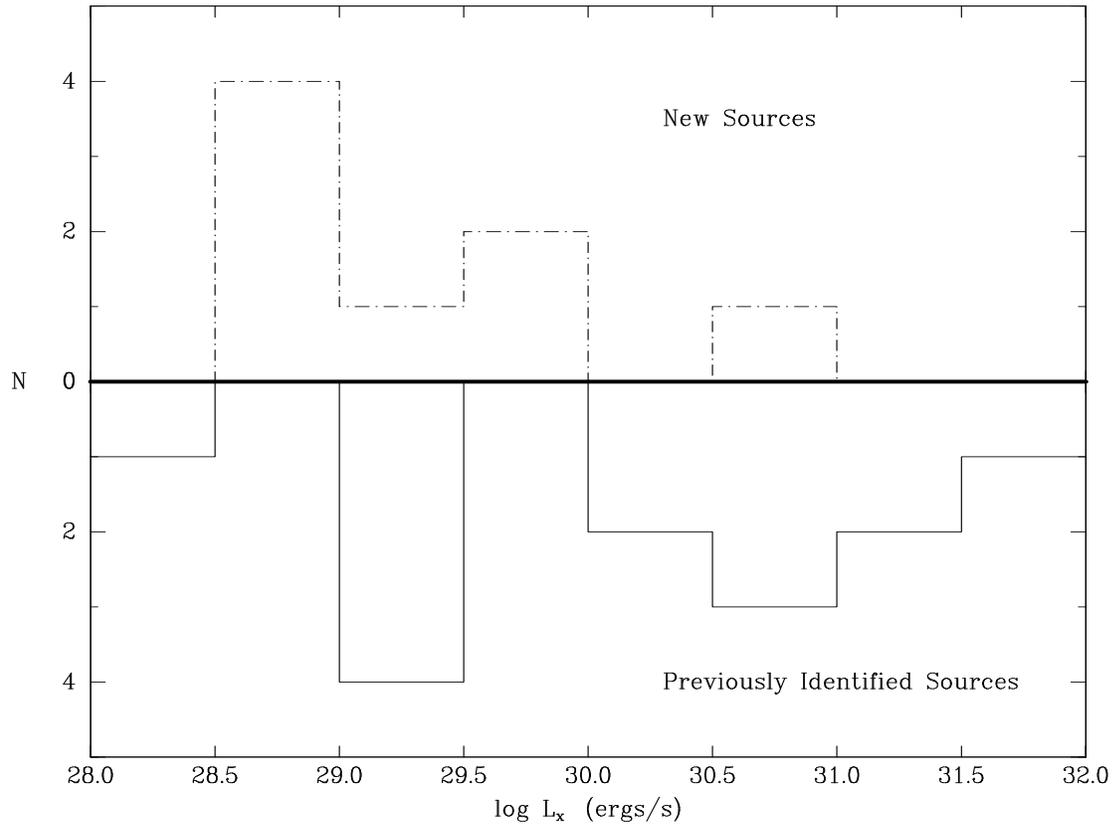}{12.0cm}{-90}{60}{60}{-250}{360}

\caption{The distribution of x-ray luminosities for the previously identified
members of the L1495E population compared with the x-ray luminosity
distribution of the newly identified cloud population.}

\end{figure}
\begin{figure}
\plotfiddle{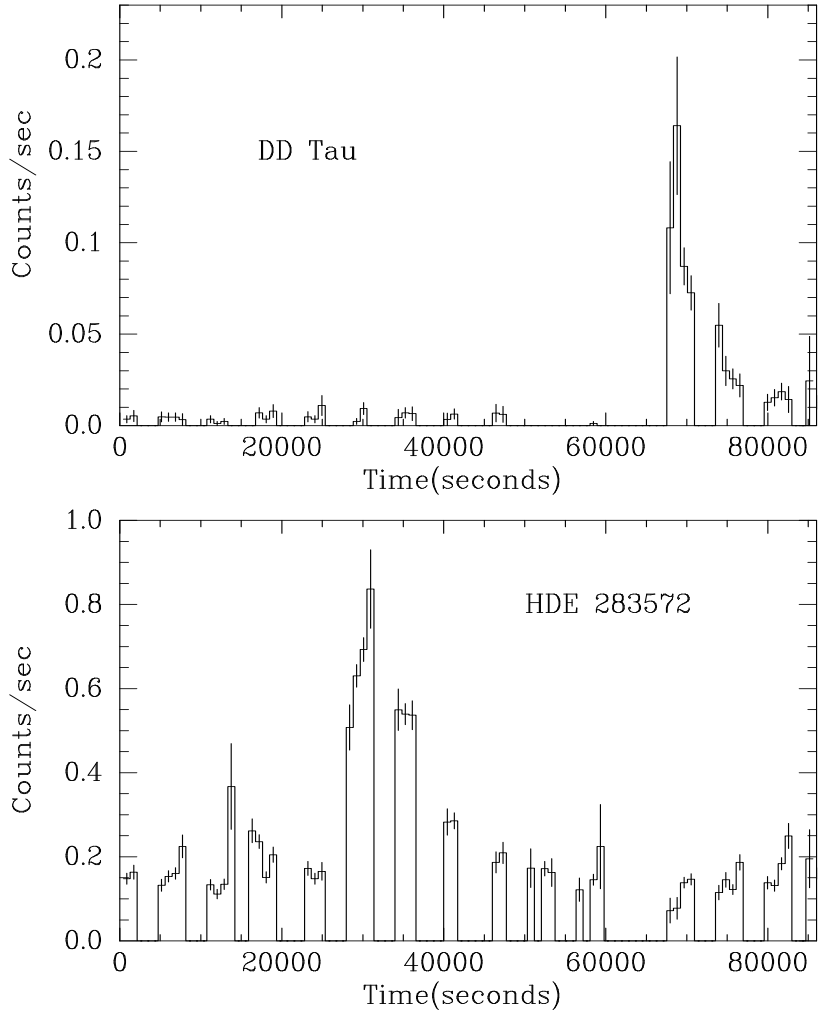}{16.0cm}{-90}{70}{70}{-275}{400}

\caption{Plots of the x-ray count rate as a function of time from the
beginning of the observation for a) DD Tau and b) for HDE 283572 binned
in 860 second intervals. Both the low level of the flux prior to the
flare and the high flare emission are obvious in DD Tau. Given the
rotation period of HDE 283572 (134000s), it is less clear whether the
event seen in the observation of this star is a small flare or is
related to the rotation of an active region on the surface to the front
face of the star.}

\end{figure}
\begin{figure}
\plotfiddle{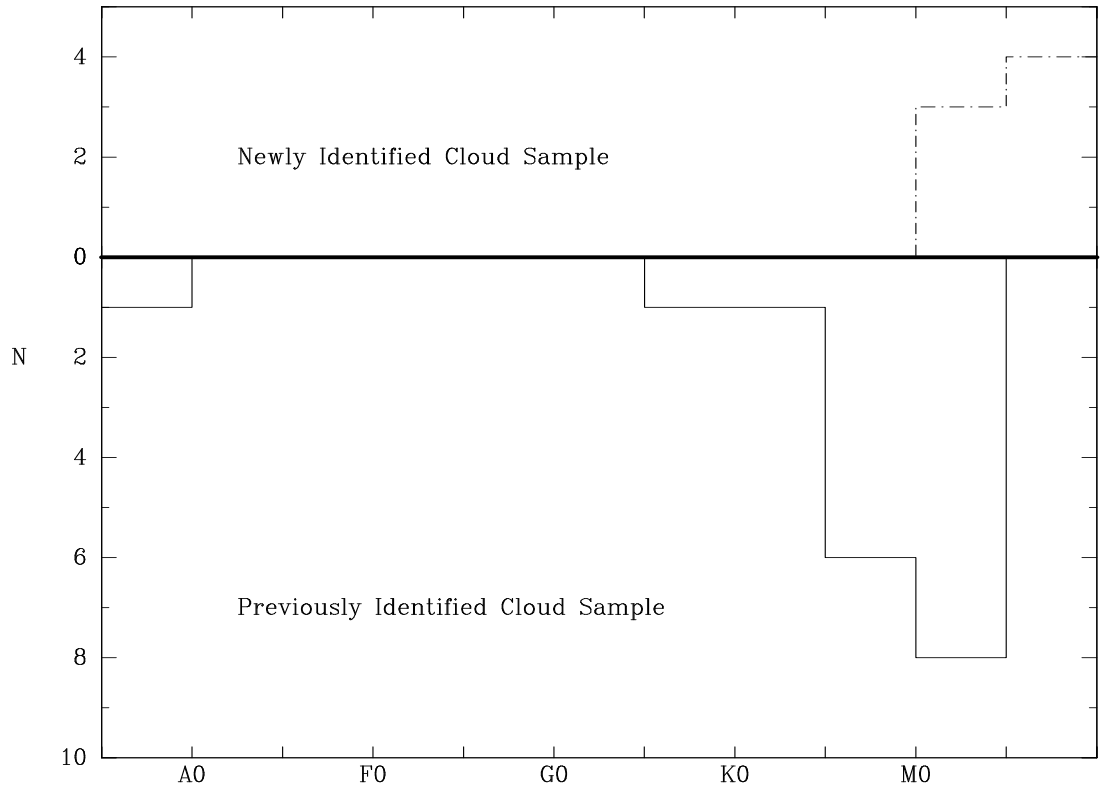}{12.0cm}{-90}{60}{60}{-250}{360}

\caption{The distribution of spectral types for the newly identified
members of the L1495E cloud population compared with the that for the
previously identified cloud population.}

\end{figure}
\begin{figure}
\plotfiddle{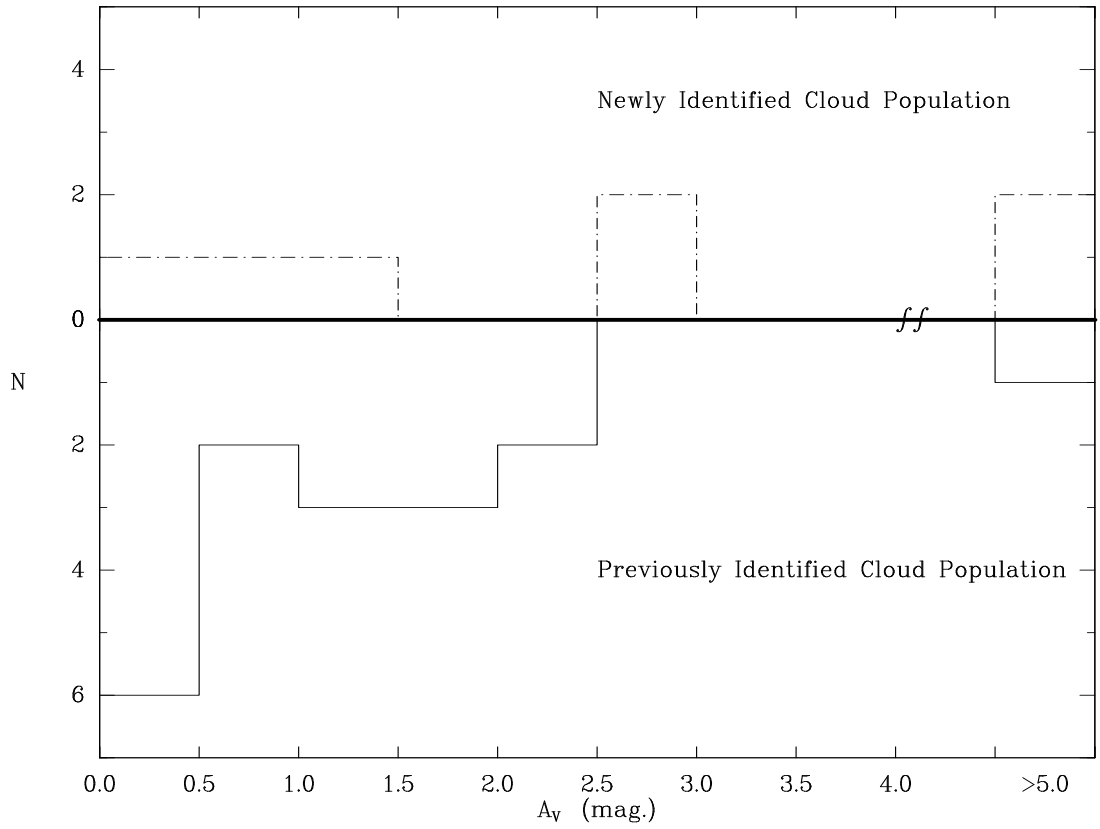}{12.0cm}{-90}{60}{60}{-250}{360}

\caption{The distribution of extinction values for the newly identified
members of the L1495E cloud population compared with the that for the
previously identified cloud population.}

\end{figure}
\begin{figure}
\plotfiddle{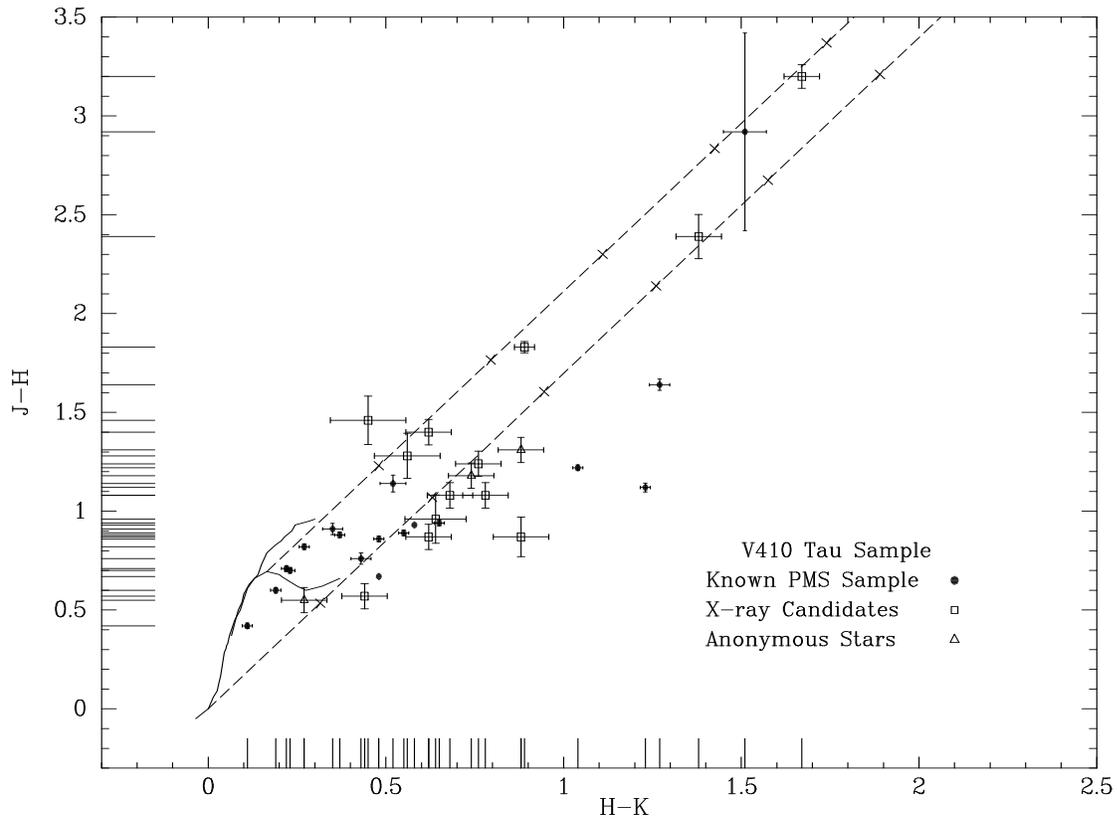}{12.0cm}{-90}{60}{60}{-250}{360}

\caption{An infrared color-color diagram for the stars in the L1495E
cloud. The filled circle show the previously identified population of
L1495E. The open squares show the candidates for identification of the
x-ray sources, and the triangles show the locations of the anonymous
objects.}

\end{figure}
\begin{figure}
\plotfiddle{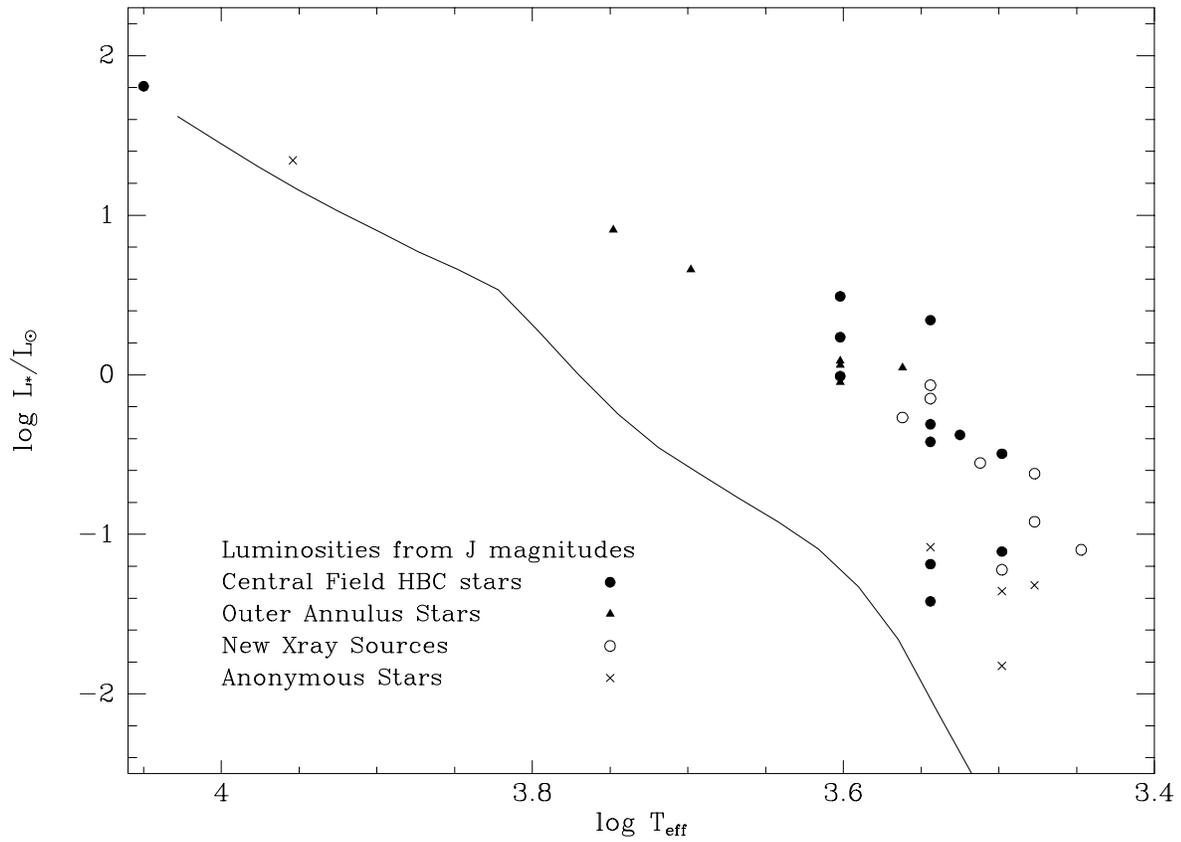}{12.0cm}{-90}{60}{60}{-250}{360}

\caption{An HR Diagram for the stars in the L1495E cloud showing only the
locations of the objects and the $10^8$ yr isochrone of D'Antona \& Mazzitelli
(1993).}

\end{figure}
\begin{figure}
\plotfiddle{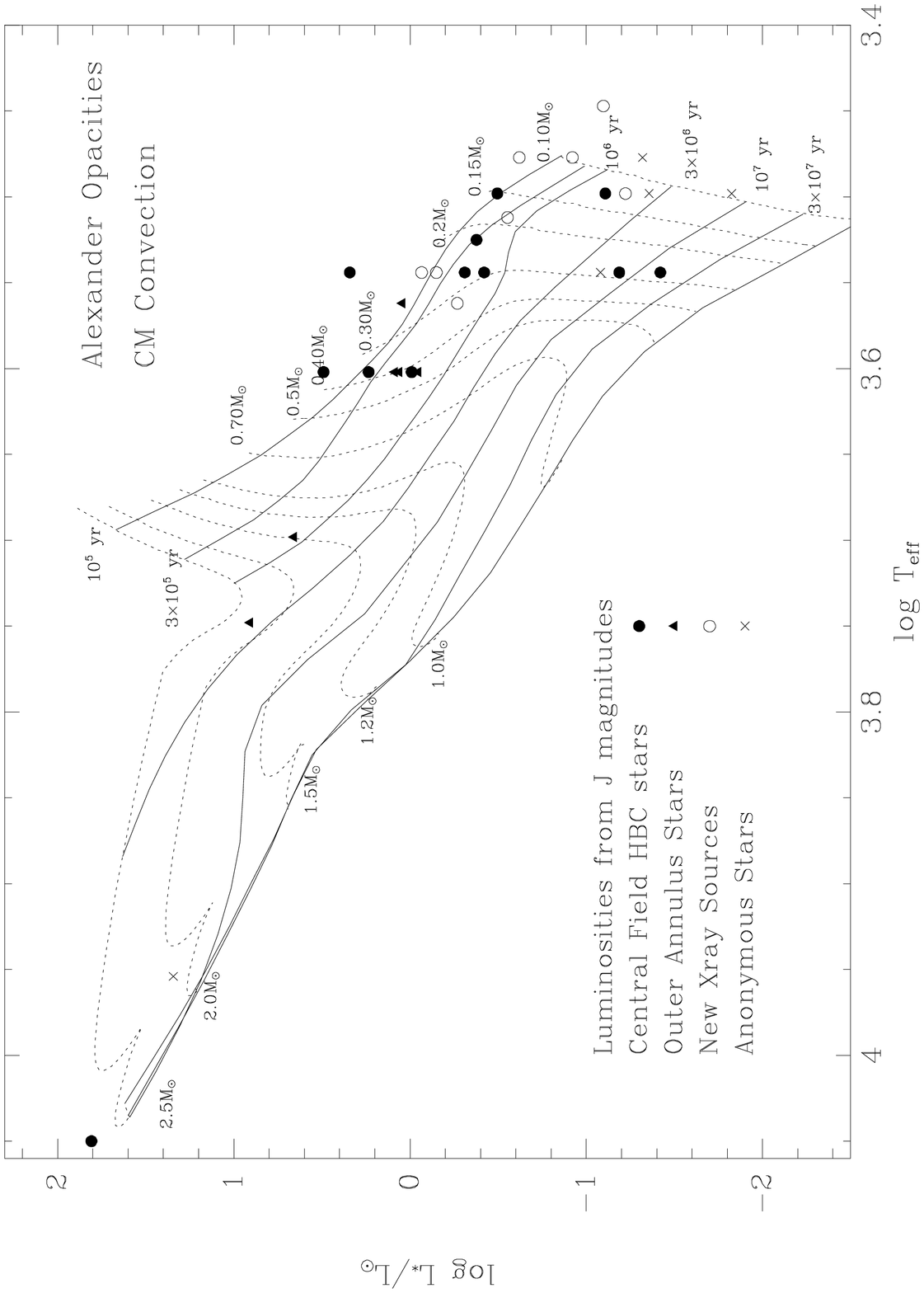}{7.5cm}{-90}{42}{42}{-180}{275}
\plotfiddle{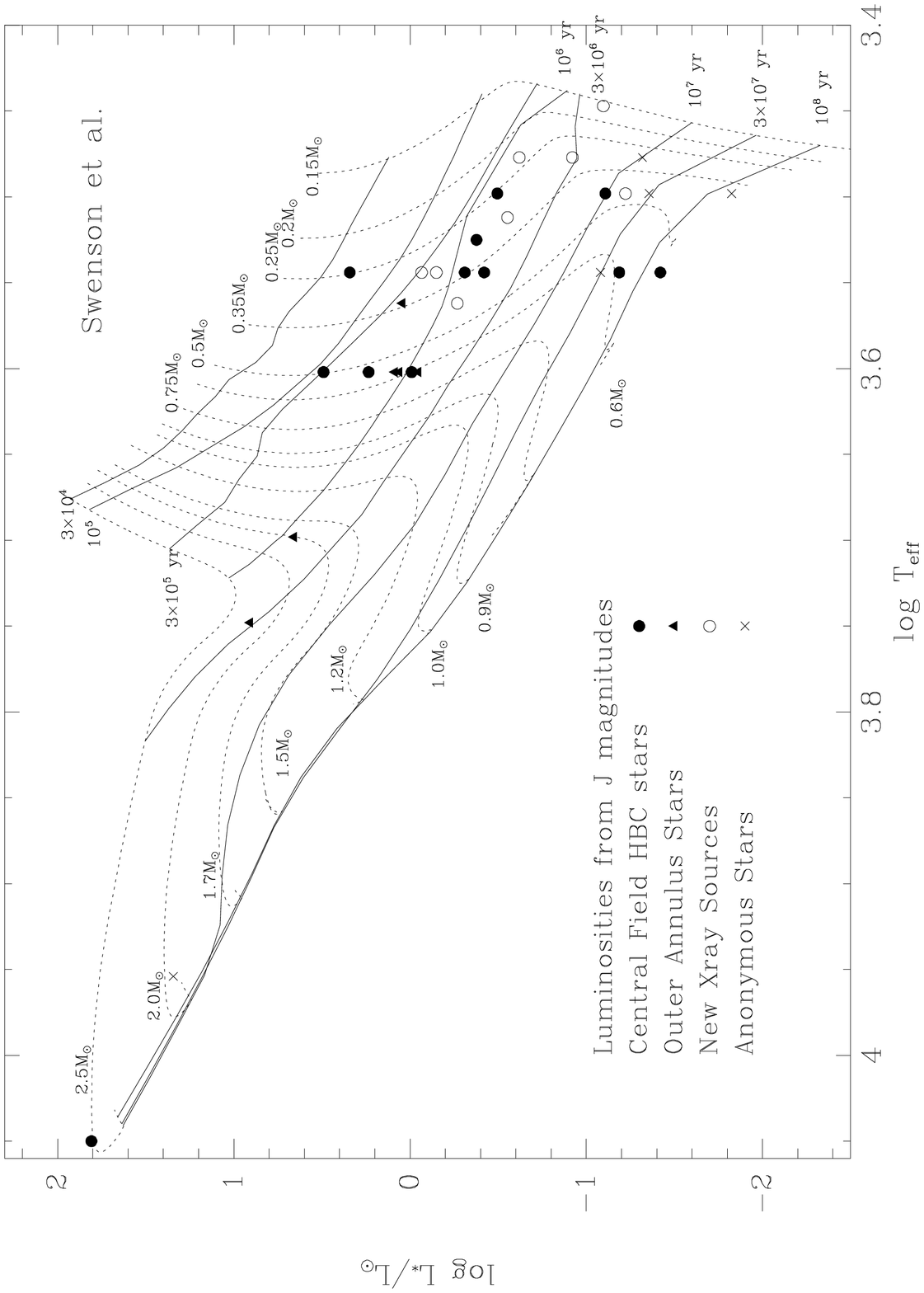}{7.5cm}{-90}{42}{42}{-180}{275}

\caption{Comparison of the stellar luminosities and effective
temperatures for the stars in L1495E with two sets of recent PMS
evolutionary tracks. a) The PMS tracks of D'Antona \& Mazzitelli (1993)
using the Alexander opacities and the convection formulation of Canuto
\& Mazzitelli, and b) the PMS tracks of Swenson et al.\ (1993). }

\end{figure}
\begin{figure}
\plotfiddle{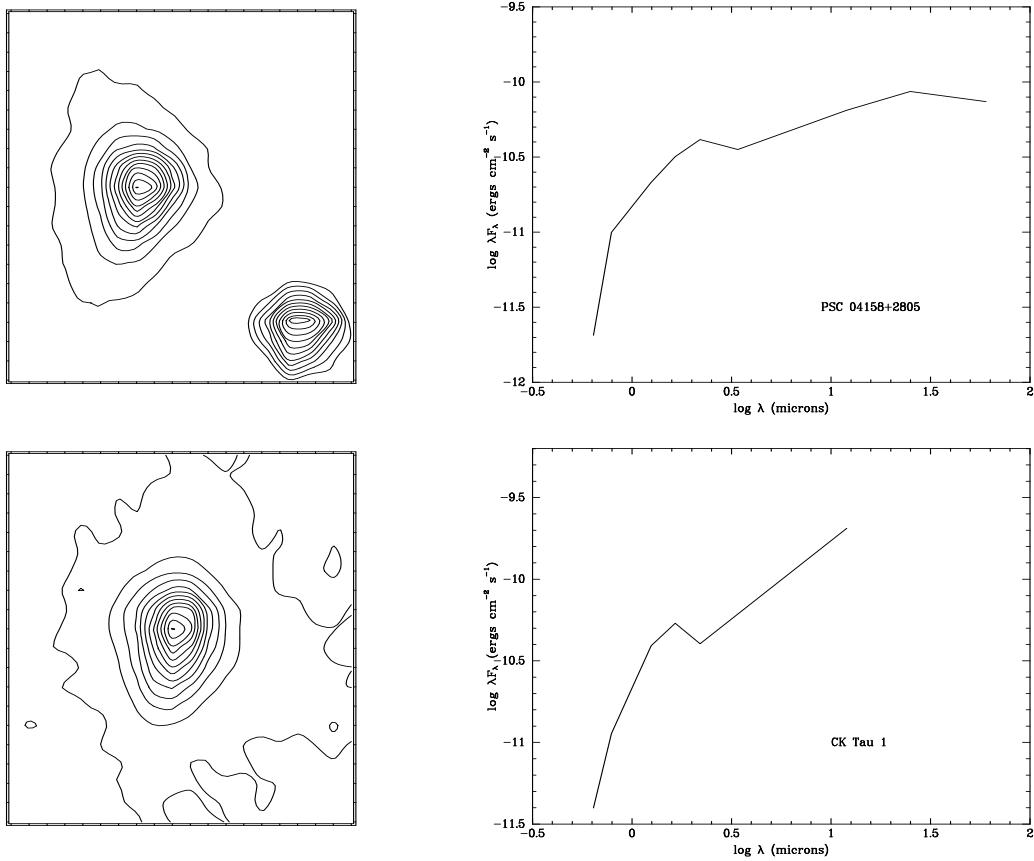}{12.0cm}{-90}{60}{60}{-250}{360}

\caption{On the left are contour plots of the I band images of CK Tau 1
and PSC 04158+2805. The plots are 20 pixels on a side and the plate
scale is 0.78\arcsec /px. In the lower right corner of the first contour
plot is shown the point spread function as exhibited by a nearby star of
similar
brightness. On the right are plots of the spectral energy
distributions of the two sources. Both objects show the secondary bump
in the spectral energy distributions characteristic of objects seen via
scattered light. }

\end{figure}
\begin{figure}
\plotfiddle{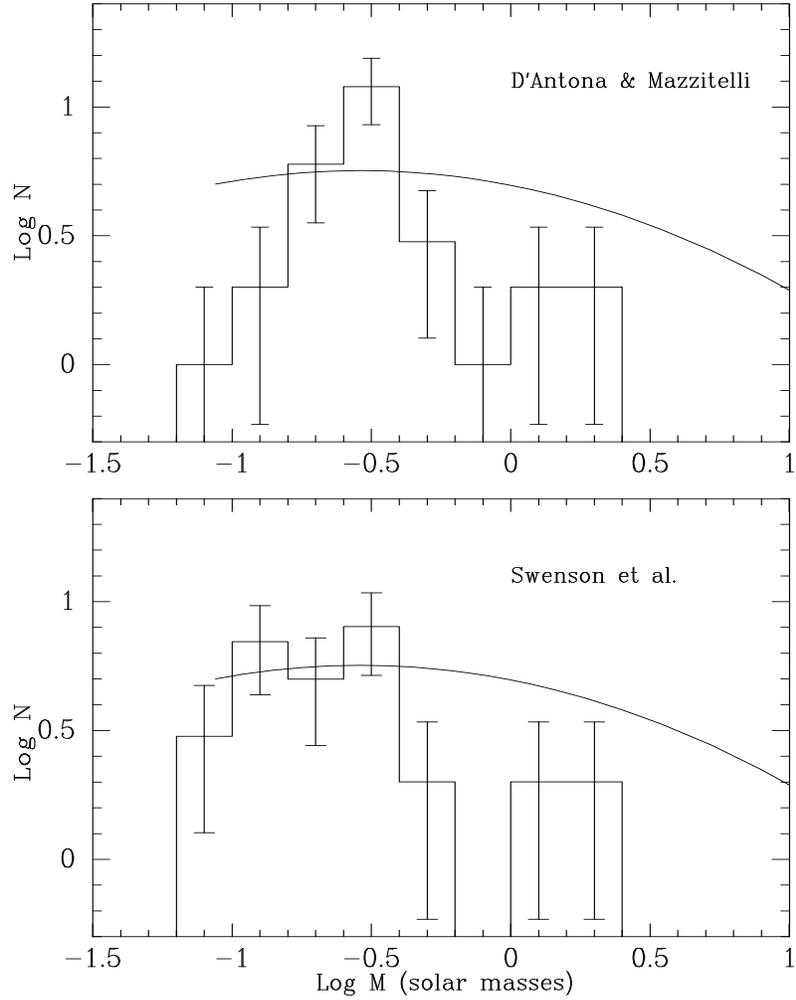}{12.0cm}{-90}{70}{70}{-275}{400}

\caption{Histograms depicting the implied IMF for the cloud PMS
population. The solid line shows the IMF derived if one uses the
Swenson et al.\ PMS tracks to infer stellar masses. The dashed line
shows the IMF derived if the D'Antona \& Mazzitelli tracks are used. The
curve shown represents the Scalo field IMF. The error bars shown
represent only $\surd N \;$ errors.}

\end{figure}

\begin{figure}
\plotfiddle{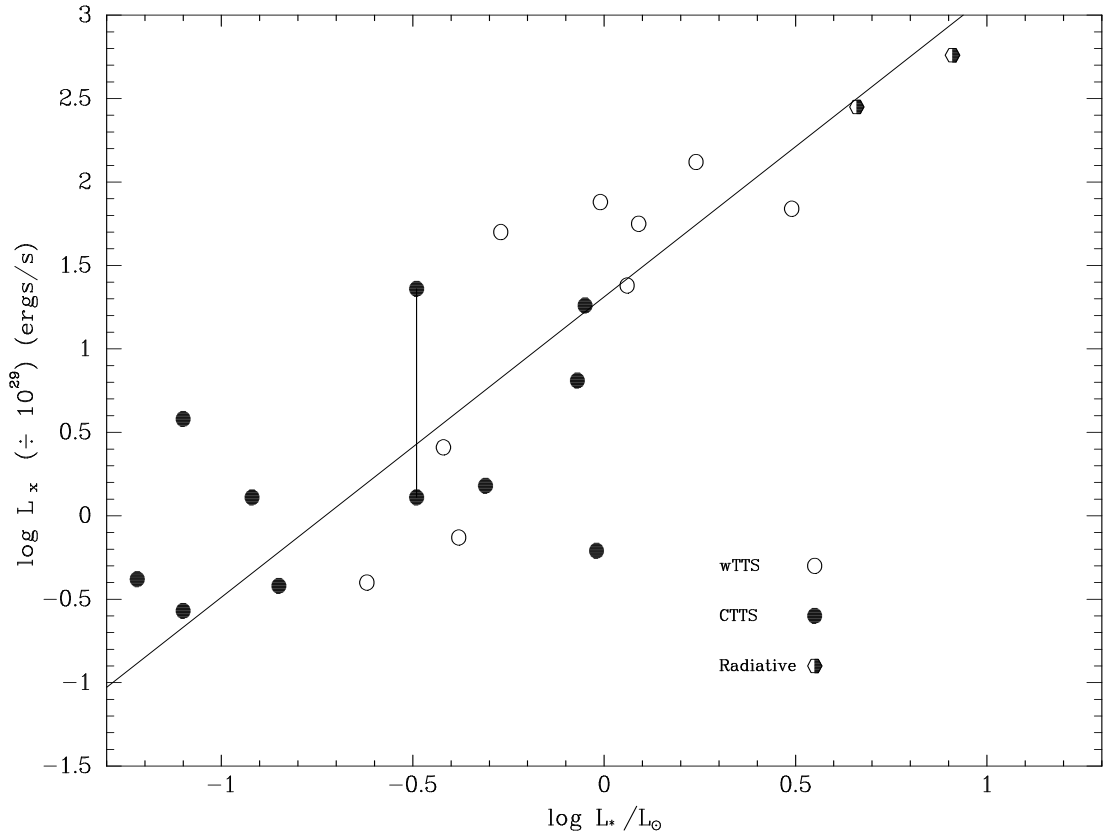}{12.0cm}{-90}{60}{60}{-250}{360}

\caption{A plot of the xray luminosity of the L1495E PMS stars versus
their stellar luminosities. The line shows the best fit to this data.
The symbols distinguish the WTTS, the CTTS and the two stars that have
evolved off their fully convective tracks and have developed radiative
cores.}

\end{figure}

\begin{figure}
\plotfiddle{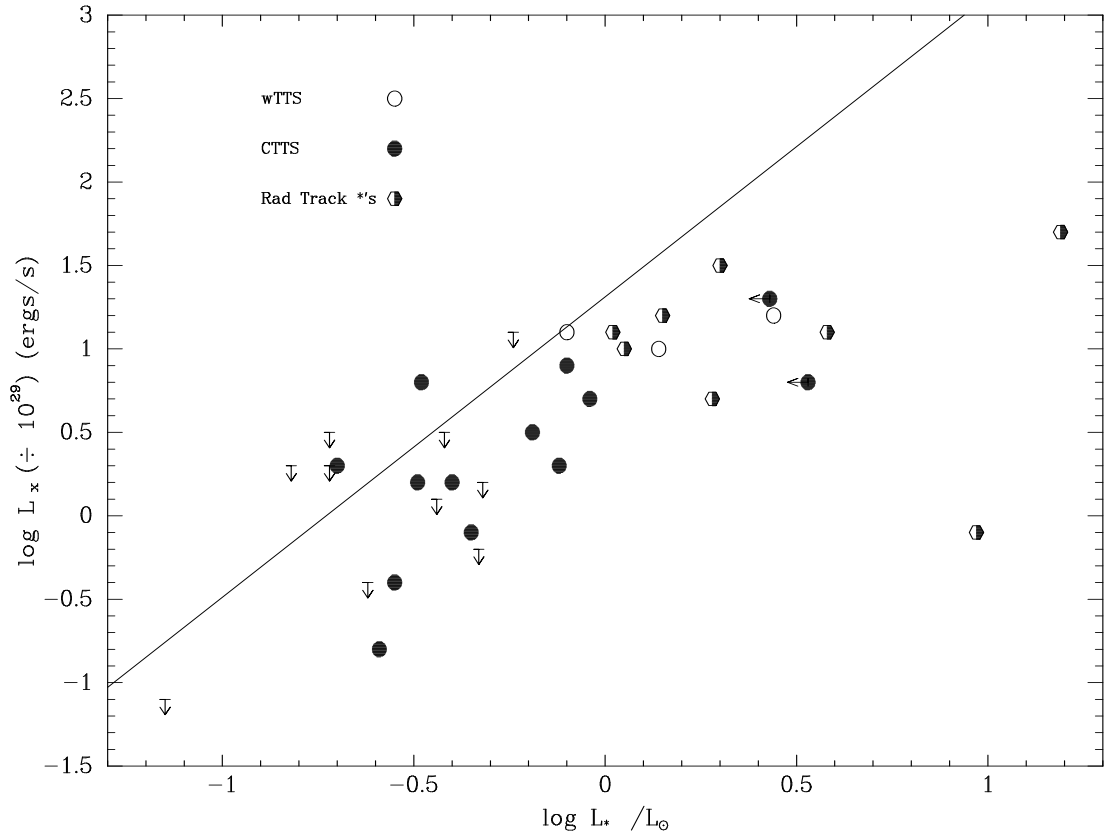}{12.0cm}{-90}{60}{60}{-250}{360}

\caption{A plot of the xray luminosity of the Chamaeleon I PMS stars versus
their stellar luminosities. The line shows the best fit to the L1495E data.
The symbols are the same as in Figure 12.}

\end{figure}

\begin{figure}
\plotfiddle{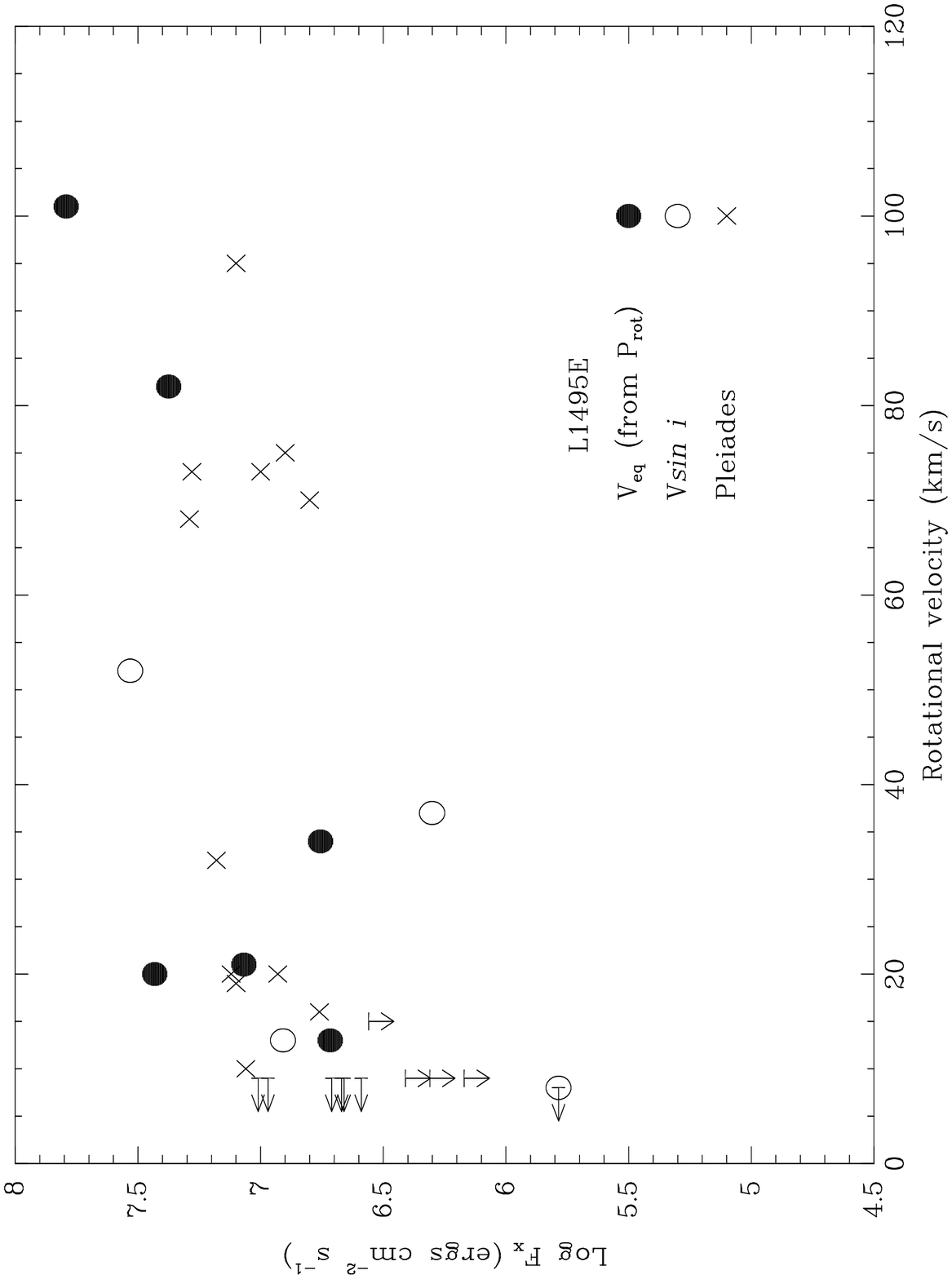}{7.5cm}{-90}{42}{42}{-180}{275}
\plotfiddle{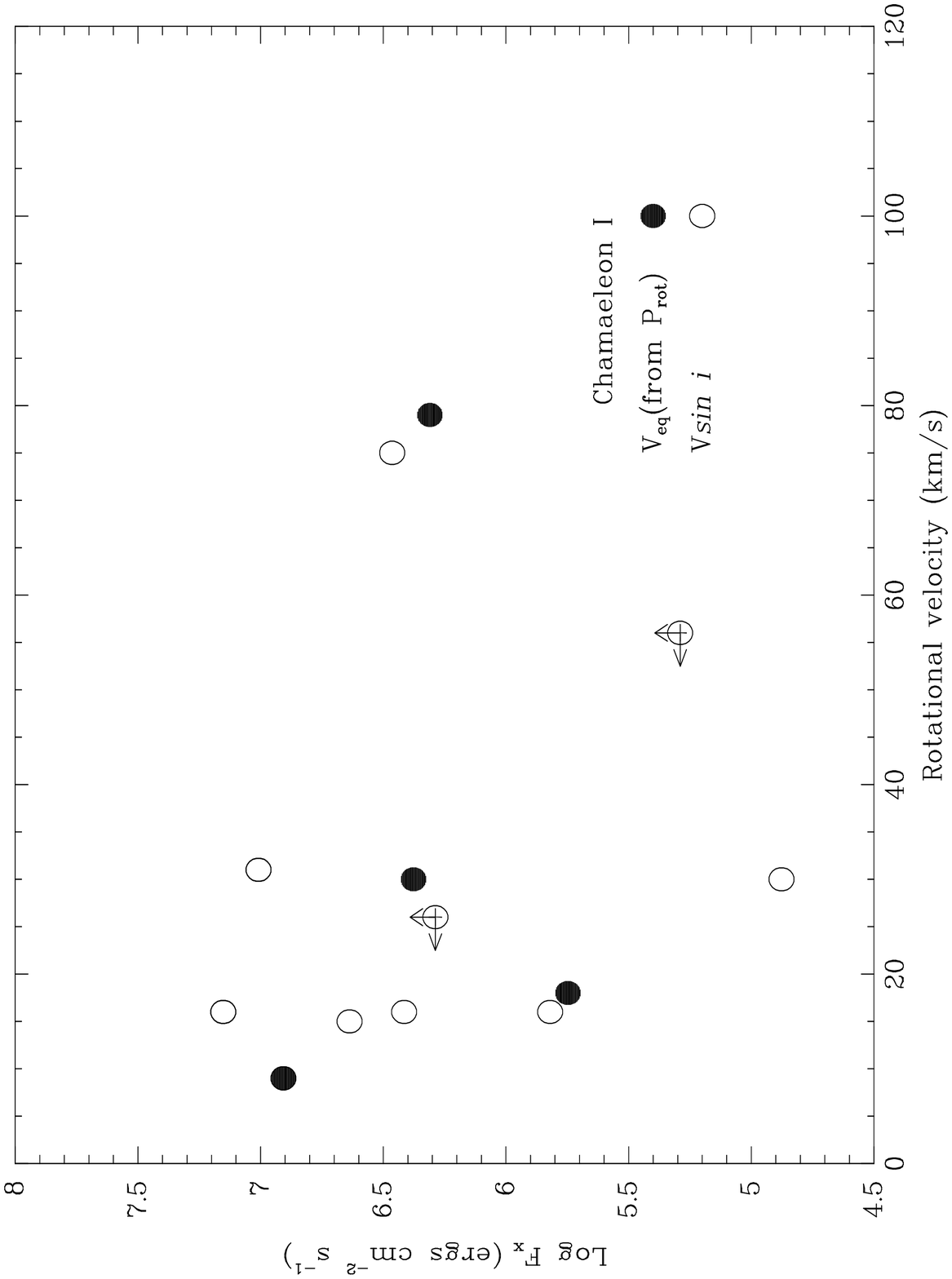}{7.5cm}{-90}{42}{42}{-180}{275}

\caption{(a)A plot of the xray surface flux versus the equatorial rotation
velocity  for the stars in L1495E and for the Pleiades (Stauffer at
al.\ (1993). The filled points represent $V_{eq}$ values derived from
observed rotation periods. The open points represent $V\sin i$ values.
The crosses show the Pleiades $V\sin i$ data.
(b) The corresponding data for the Cha~I stars. The meaning of the
symbols is the same as in the previous figure. The two points with
arrows superposed, show the direction of corrections expected in the
rotation velocity and x-ray surface flux due to the heavy veiling of
these two stars. See discussion in section 7.}

\end{figure}

\end{document}